\documentclass[12pt,preprint]{aastex}

\shorttitle{Mid-IR properties of EGS X-ray sources}
\shortauthors{Barmby et al.}

\newcommand{\mic}{~$\mu$m\ }
\newcommand{\mice}{~$\mu$m}
\newcommand{\cgs}{~erg~cm$^{-2}$~s$^{-1}$}
\newcommand{\xopt}{X--ray to optical ratio\ }

\begin{document}

\title{Mid-infrared properties of X-ray sources in the Extended Groth Strip}

\author{P. Barmby\altaffilmark{1}, 
A. Alonso-Herrero\altaffilmark{2,3}, 
J.L. Donley\altaffilmark{3},
E. Egami\altaffilmark{3},
G.G. Fazio\altaffilmark{1}, 
A. Georgakakis\altaffilmark{4},
J.-S. Huang\altaffilmark{1}, 
E.S. Laird\altaffilmark{4,5},
S. Miyazaki\altaffilmark{6},
K. Nandra\altaffilmark{4},
S.Q. Park\altaffilmark{1},
P.G. P\'erez-Gonz\'alez\altaffilmark{2}, 
G.H. Rieke\altaffilmark{3},
J.R. Rigby\altaffilmark{3},
S.P. Willner\altaffilmark{1}
}

\altaffiltext{1}{
Harvard-Smithsonian Center for Astrophysics, 60 Garden St., Mailstop 65, Cambridge, MA 02138}
\altaffiltext{2}{
Departamento de Astrof\'{\i}sica Molecular e Infrarroja, Instituto de Estructura de la Materia, CSIC, 
Serrano 113bis \& 121, 28006 Madrid, Spain}
\altaffiltext{3}{
Steward Observatory, The University of Arizona, 933 N. Cherry St., Tucson, AZ 85721}
\altaffiltext{4}{
Astrophysics Group, Imperial College London, Blackett Laboratory, Prince Consort Rd., London SW7 2AW, UK}
\altaffiltext{5}{Now at UCO/Lick Observatory, Department of Astronomy and Astrophysics, 
University of California, Santa Cruz, CA 95064}
\altaffiltext{6}{
Subaru Telescope, National Astronomical Observatory of Japan, Hilo, HI 96720, USA}

\begin{abstract}
Mid-infrared observations of Active Galactic Nuclei (AGN) are 
important for understanding of the physical conditions
around the central accretion engines.
{\it Chandra} and {\it XMM-Newton} X--ray observations of 
a 300 arcmin$^2$ region in the Extended Groth Strip are used to
select a sample of $\sim 150$ AGN.
The {\it Spitzer} instruments IRAC and MIPS detect 68--80\% of these 
sources, which show a wide range of mid-infrared properties. About 
40\% of the sources have red power-law spectral
energy distributions ($f_{\nu} \propto {\nu}^{\alpha}$, $\alpha<0$)
in the $3.6-8$\mic IRAC bands. In these sources
the central engine dominates the emission at both X--ray and IR
wavelengths. Another 40\% of the sources have blue mid-IR spectral energy 
distributions ($\alpha>0$) with their infrared emission
dominated by the host galaxy; the remaining 20\% are not well-fit by
a power law. Published IRAC color criteria for AGN select most of the red sources, but only some
of the blue sources. As with all other known methods, selecting AGN
with mid-IR colors will not produce a sample
that is simultaneously complete and reliable.
The IRAC SED type does not directly correspond to 
X--ray spectral type (hard/soft).
The mid-IR properties of X--ray-detected Lyman-break, radio, submillimeter,
and optically-faint sources vary widely and, for the most part, are not
distinct from those of the general X--ray/infrared source population.
X--ray sources emit 6--11\% of the integrated mid--IR light,
making them significant contributors to the cosmic infrared background.
\end{abstract}


\section{Introduction}

Understanding the nature of active galactic nuclei (AGN) and the galaxies that
host them is important for such diverse goals as pinpointing the
sources of the cosmic X--ray and infrared backgrounds and deriving
the star formation history of the universe.
Multi-wavelength surveys are particularly important for
the study of AGN because their appearance in different
wavelength regimes can be quite different. Selecting 
AGN at one particular wavelength is no guarantee of 
a complete sample. X--ray selection has the advantages of 
being reasonably efficient and reliable \citep{mush04} but may
miss some obscured sources \citep{pet05}.
Infrared and radio observations can identify AGN missed in the X--ray \citep{donley05,aah05}
and, for X--ray-selected AGN, can help to distinguish
between the different AGN types. Infrared data are also needed to constrain the fraction
of the cosmic infrared background (CIRB) emitted by AGN.

The limited spatial resolution of previous generations of X--ray and infrared 
observatories made cross-identification between wavelengths difficult.
Recent work has benefited from the much smaller point spread
functions of the {\it Chandra X--ray Observatory}, {\it XMM-Newton},
the {\it Infrared Space Observatory (ISO)}, and the {\it Spitzer Space Telescope}.
\citet{fadda02} combined {\it Chandra}, {\it XMM}, and {\it ISO}
data from the Lockman Hole and Hubble Deep Field North to conclude that AGN 
contribute $15\pm5$\%  of the CIRB at 15\mice.
Work with {\it Spitzer} data to date has concentrated on the properties of 
the X--ray source population.
\citet{rigby04} studied Chandra Deep Field South sources detected at hard X--ray and 24\mic 
wavelengths and  found, surprisingly, that X--ray-hard AGN are not infrared-brighter
(as would be expected if they were embedded in the obscuring matter).
\citet{aah04} found that similarly-selected Lockman Hole sources 
exhibit a variety of optical/IR spectral types. About half of their sources
had spectral energy distributions (SEDs) dominated by stellar emission or showed significant obscuration.
\citet{fra05} found a similar mix of spectral types among ELAIS-N1 {\it Chandra}/{\it Spitzer}
sources and concluded that about 10--15\% of 24\mic sources are dominated by an AGN.

This paper combines {\it Spitzer}, {\it Chandra}, and {\it XMM} observations 
to understand the mid-infrared properties of the X--ray sources in
the `Extended Groth Strip' (EGS). The X--ray and infrared observations in this region are
intermediate in depth and area between GOODS \citep{goods} and the shallower 
NOAO Deep-Wide Field \citep{ndwfs,irac_shallow} and SWIRE \citep{swire} surveys. 
As such, the EGS provides a valuable probe of the properties of AGN at
intermediate fluxes and an additional measure of the cosmic variance of
those properties.
The extensive multi-wavelength observational data available for
the EGS, particularly its spectroscopic redshift survey, 
should eventually produce an extremely thorough characterization of
the X--ray sources and allow comparisons and cross-identifications
with other classes of galaxies. In this paper we focus on combining
mid-IR and X--ray data to study the properties of AGN and their host galaxies.
Identifications of X--ray sources in the {\it Chandra} deep fields \citep{horn03, bar03}
suggests that almost all of the EGS X--ray sources should be AGN.
The EGS X--ray sources' median redshift is expected to be $z\sim1$ with no strong dependence 
on X--ray flux \citep{bar05}.

\section{Observational Data}

The Extended Groth Strip (EGS) is a high ecliptic- and Galactic-latitude
field which extends the original {\it Hubble Space Telescope}
`Groth-Westphal strip' \citep{gws} to a size of 2\arcdeg $\times$ 15\arcmin.
The EGS is one of the target fields for the DEEP2 redshift survey \citep{deep2},
and has extensive optical and near-infrared ground-based imaging \citep{deep2opt,deep2ir}.
An approximately $20\arcmin \times 20\arcmin$ area in its southwest corner
(the `14-hour field') contains several optical redshift surveys \citep{cfrs,ccs03},
submillimeter \citep{egs_scuba, webb03} and ISO observations \citep{egs_iso},
a deep radio survey at 5~GHz \citep{egs_vla91}, and X--ray observations with both
{\it XMM-Newton} \citep{was03,miy04}, and {\it Chandra} \citep{egs_cxo}.
Observations of the full EGS at ultraviolet and radio wavelengths are in progress,
as are additional {\it Chandra} observations which will eventually cover the entire
EGS to a depth of 200~ks.

The {\it Spitzer} data used here were obtained as part of program 8, the 
``IRAC Deep Survey'' with observing time contributed by Guaranteed
Time Observers G.~Fazio, G.~Rieke, and E.~Wright. The IRAC observations for this
program were obtained at two epochs, 2003 December and 2004 June--July and
cover a $2\arcdeg \times 10\arcmin$ area to a depth of $\sim2.7$~h per sky position. 
Observations with the MIPS instrument were begun in 2004 January but were not completed
because of a spacecraft safing event. The full-depth MIPS observations, covering approximately
the same area as the IRAC data to a depth of 1200~s per sky position at 24 \mice, were completed 
in 2004 June.\footnote{The analysis
of a subset of the EGS X--ray sources presented by \citet{aah04} was based on only the 2004 January 
MIPS and 2003 December  IRAC data.} The IRAC and MIPS data reduction was similar to that
described by \citet{irac_lh} and \citet{egami04}. The 
$5\sigma$ limiting point-source flux densities of the data are (0.9, 0.9, 6.3, 5.8, 83) $\mu$Jy 
[limiting AB magnitudes of 24.0, 24.0, 21.9, 22.0, 19.1] at (3.6, 4.5, 5.8, 8.0, 24)\mice.\footnote{
Unless stated otherwise, all IRAC magnitudes and colors as well as IRAC-MIPS colors
in this paper are on the AB system. $[\lambda]$ signifies a magnitude in the $\lambda-\mu$m band.}
The 24\mic source identification is 80\% complete at the 83~$\mu$Jy limit 
(about a factor of 1.5 above the confusion limit), while the completeness of
the 3.6 and 4.5\mic data are substantially affected by source confusion at faint flux densities
\citep{irac_numcts}. For comparison to the X--ray sources,
we generated two catalogs from the full-area EGS data. 
The IRAC-only comparison sample is selected at 8\mic and contains
$\sim18000$ sources, with the area overlapping the X-ray surveys containing about 4400 8\mic sources.
The IRAC/MIPS comparison sample includes all sources detected at 24\mic and in the IRAC bands
and contains $\sim 4700$ sources in the $2\arcdeg \times 10\arcmin$ area.

\begin{figure}
\plotone{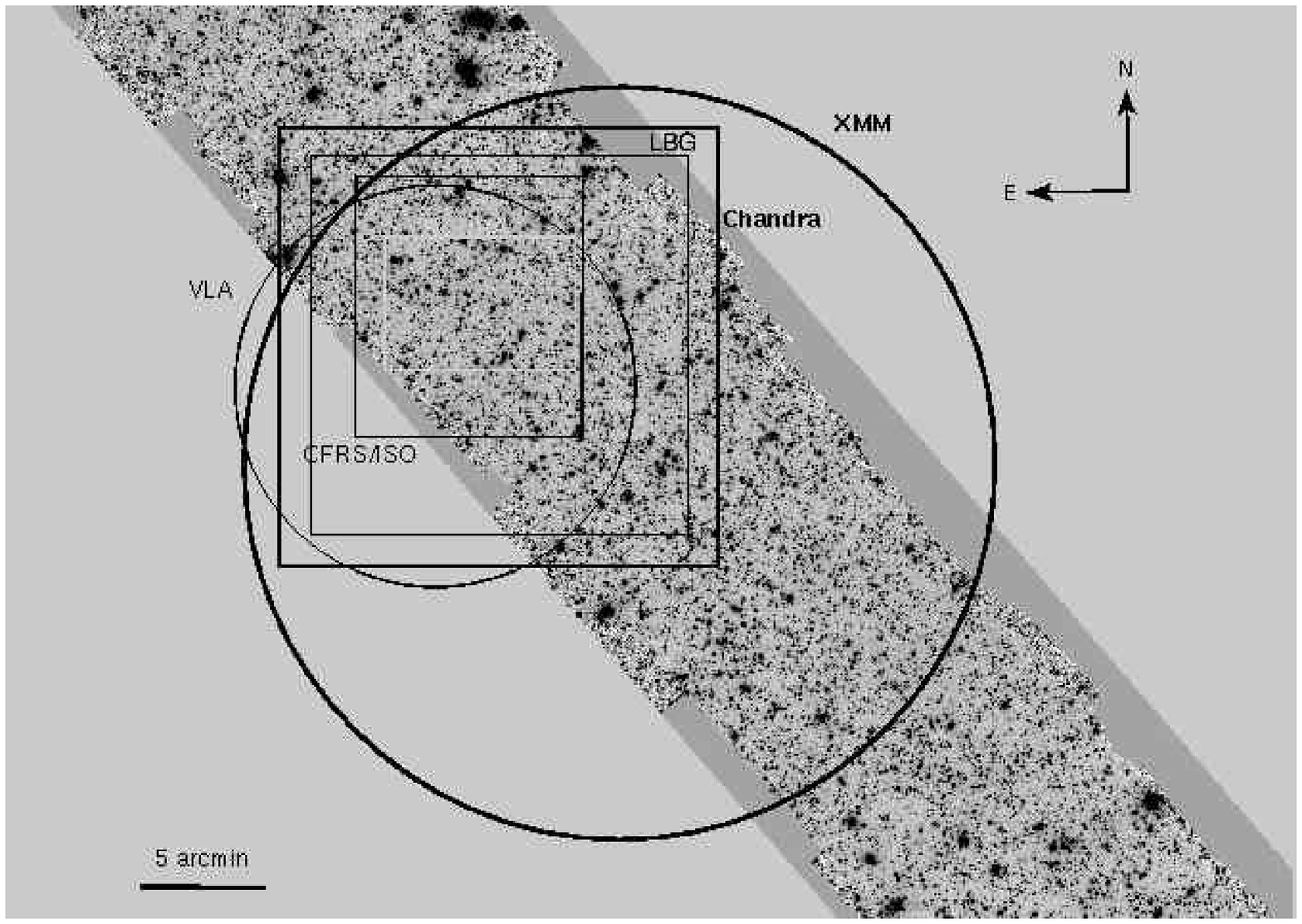}
\caption{EGS survey regions, superimposed on the IRAC 3.6\mic image.
Bold black lines are the XMM (circle) and Chandra (square) pointings.
Other surveys include the Lyman-break galaxy survey of \citet{ccs03},
the Canada-France Redshift Survey and ISO observations \citep{cfrs, egs_iso},
the VLA survey of \citet{egs_vla91} and SCUBA survey of \citet{egs_scuba}.
\label{fields}}
\end{figure}

The details of the X-ray observations and source lists used here are described
elsewhere \citep{was03,egs_cxo}. Briefly, the {\it XMM-Newton} data consist
of a 56~ks exposure obtained in 2000 July  (ObsID 127921001, PI: Griffiths), centered on the position
$14^{\mathrm h}17^{\mathrm m}12$\fs0, $52\arcdeg25\arcmin00$\fs0 (J2000).
The catalog produced by \citet{was03} contains 154 sources detected in either/both of the
soft (0.5--2~keV) or hard (2--10~keV) bands, with a limiting 0.5--10~keV
flux of $2\times 10^{-15}$\cgs. The expected number of spurious sources is $\sim 3$.
The {\it Chandra} data consist of 200~ks of observations
with ACIS-I obtained in 2002 August (ObsIDs 3305, 4357, 4365, PI: Nandra), 
centered on position $14^{\mathrm h}17^{\mathrm m}43$\fs6, $52\arcdeg28\arcmin41\arcsec$ (J2000).
The catalog produced by 
\citet{egs_cxo} contains 158 sources detected in the full (0.5--7~keV)
and/or soft (0.5--2~keV) X--ray bands; the expected number of spurious sources is 1.8.
The limiting full-band flux, converted to the standard 0.5--10~keV band,
is $3.5\times 10^{-16}$\cgs. The fields observed by the two X--ray instruments partially
overlap and extend outside the region observed by {\it Spitzer}; 
the area in common between the {\it Spitzer} and {\it Chandra}
surveys is about 186~arcmin$^2$, with the {\it XMM}-only/{\it Spitzer} area adding another
114~arcmin$^2$.
Figure~\ref{fields} shows the geometry of the various surveys.

We combined the two X--ray catalogs to produce a final list of X--ray sources.
Sources from the {\it Chandra} and {\it XMM} catalogs were matched
using the X-ray positions given by \citet{egs_cxo} and the optical counterpart positions
given by \citet{was03} (these should be more precise than the {\it XMM} X--ray positions). 
The matching radius between the two catalogs was 5\arcsec; we expect
few false matches because the source density is low.
A total of 152 X--ray sources are within the boundaries of the {\it Spitzer} observations:
39 {\it XMM}-only sources (6 within the {\it XMM-Chandra} overlap area), 72 {\it Chandra}-only
sources (all within the overlap area), and 42 sources in common between the two X--ray
lists. (The large number of {\it Chandra}-only sources within the overlap area
is presumably due to the much fainter flux limit of the {\it Chandra} observations.)
For the sources detected by both {\it Chandra} and {\it XMM}, our subsequent analysis
uses the X--ray positions and fluxes given by \citet{egs_cxo}.
To allow direct comparison between {\it Chandra} and {\it XMM} sources, 
we computed hardness ratios from the tabulated fluxes%
\footnote{Hardness ratios are defined as HR=$(H-S)/(H+S)$ where $H$ and $S$ are the
fluxes in the 2--10~keV hard band and the 0.5--2~keV soft band, respectively.}
rather than from image counts as is often done.

Most of the effort to date in classifying X--ray sources has involved 
optical identifications and spectroscopy. Although the focus of this work is
the X--ray sources' mid-infrared properties, we use deep $R$-band images taken with the Suprime 
camera \citep{suprime} on the Subaru telescope
for comparison with other work and to identify any optically-faint, infrared-bright
sources. The four 36\arcmin$\times 28$\arcmin\ %
images cover the EGS completely to a depth of $R\approx26.5$ with seeing FWHM
0\farcs6--0\farcs8.
The optical counterparts of the {\it Chandra} sources are discussed in detail
elsewhere \citep{geo05}.
The optical counterparts of the {\it XMM} sources within the original HST/WFPC2 
`Groth Strip' are discussed by \citet{miy04}.

\subsection{Cross-matching and Photometry\label{sec:phot}}

With the X--ray catalog positions  as the input to  
the IRAF/APPHOT {\sc center} and {\sc phot} routines, we searched for
$>5\sigma$ 3.6\mic sources within a 2\farcs5 radius. 
Only a handful of X--ray sources had multiple sources within the search radius, and
we do not consider these as secure detections. Of the 152 X--ray sources, 
138 (91\%) have secure IRAC detections; all are  detected in all four IRAC bands.
Detection of MIPS counterparts was performed by 
searching for the nearest detectable source within 3\arcsec.
All of the 104 X--ray sources (68\% of the total) detected by MIPS at 24\mic 
are also detected by IRAC. 
Because the IRAC 3.6\mic images are crowded, some
false X--ray/infrared matches are to be expected. To quantify this effect,
we shuffled the X-ray catalog to produce random coordinates within the
X--ray/IR overlap region and re-performed the centering and photometry
procedure. About 18\% of the random ``sources'' were detected
in the 3.6\mice-band image; less than half of these (about 8\% of the total)
were detected in the less-crowded 8.0\mice-band image. We conclude that the false-match
probability is likely to be about 10\%.

\begin{figure}
\plotone{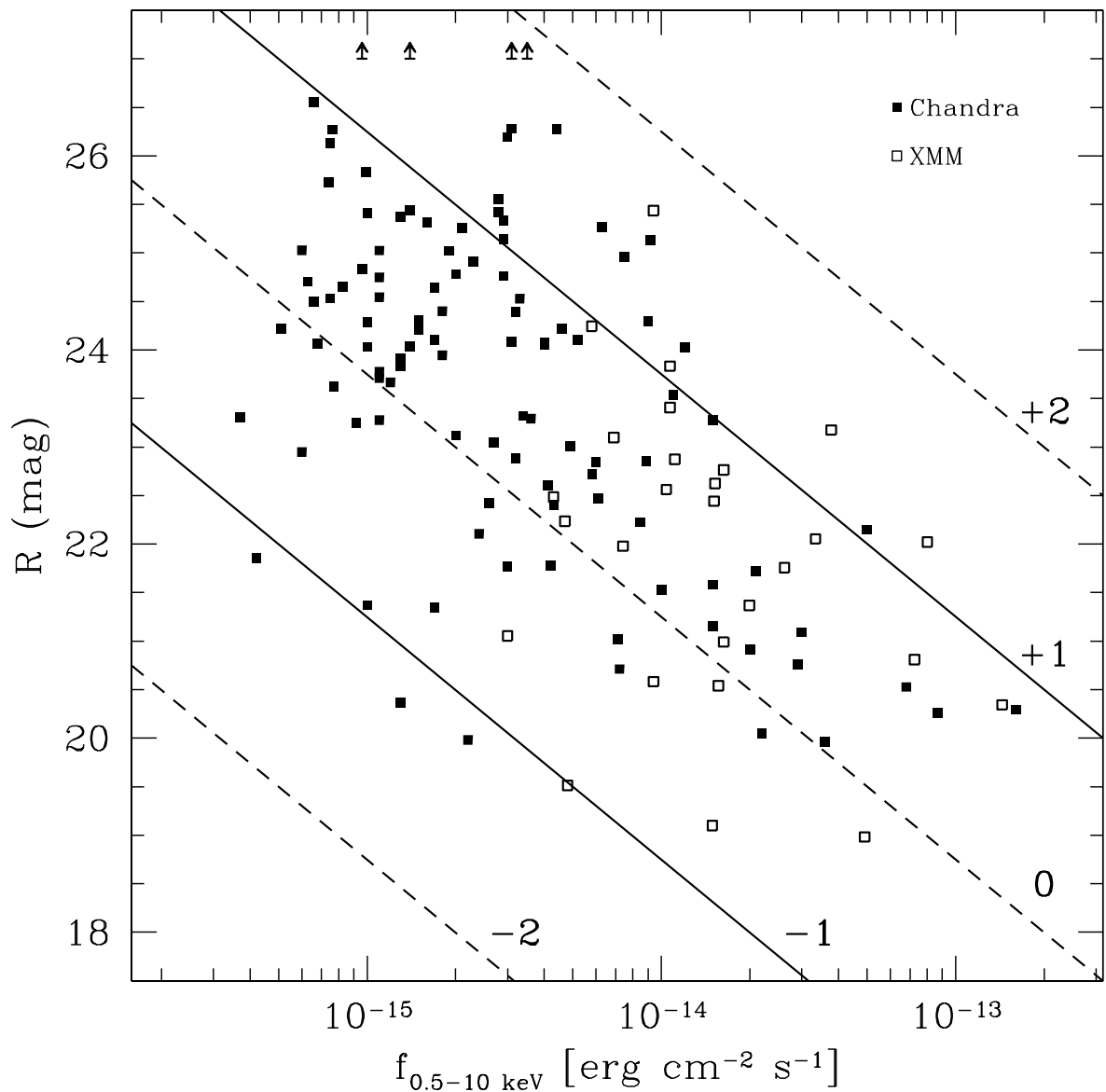}
\caption{X--ray flux versus optical magnitude for EGS X--ray sources.
Filled (open) squares indicate sources detected with {\it Chandra} ({\it XMM}).
Diagonal lines have constant values of $\log (f_X/f_R) = \log(f_X) + 5.50 + R/2.5$ 
\citep{horn03}, labeled at right.
\label{fx_opt}}
\end{figure}

Detection of the optical counterparts was performed by using the IRAF
{\sc apphot} routine with the aperture centered on the IRAC position.
Of the fourteen IRAC-unmatched sources X--ray sources, six are {\it Chandra}
sources which do not appear to have optical counterparts, and seven are
{\it XMM-Newton} sources for which an optical counterpart was not identified
by \citet{was03}. All but one of the {\it XMM-Newton} sources lacking IRAC 
counterparts have very faint optical point sources within 5\arcsec\ of the X--ray position.
Since the focus of this work is on the infrared properties 
we do not discuss these sources further. 
There are four {\it Chandra} sources that have IRAC and MIPS counterparts but lack
optical counterparts; these sources are discussed further 
in Section~\ref{sec:subset}. Figure~\ref{fx_opt} shows the \xopt for all of the 
IR-detected sources. As expected given the X--ray survey depth, most have $\log (f_X/f_R)>-1$, 
indicating that they are likely to be AGN and not normal galaxies.

IRAC photometry was performed using the source positions determined from the 3.6\mic image. 
We measured flux densities in 1\farcs8-radius apertures and corrected to
total magnitudes in the standard 12\farcs2-radius calibration aperture using 
a `mosaic PSF' for each channel derived from the IRAC images.%
\footnote{The corrections used were 0.55, 0.59, 0.73 and 0.84 magnitudes in the 4 IRAC bands.}
Photometry on the MIPS 24\mic image was done using PSF-fitting, as described by \citet{pg05}.
Photometry on the Subaru $R$-band image was measured in 1\farcs0-radius apertures, 
aperture-corrected to $r=$2\farcs0. Photometric uncertainties used for the aperture
photometry are those returned by {\sc phot}, with the IRAC uncertainties multiplied
by two to account for correlated noise in the backgrounds.
Table~\ref{data_tab} gives the infrared properties of the detected X--ray sources.
A handful of the X--ray sources are also matched with sources in the various
other surveys which overlap the region. These are marked in Table~\ref{data_tab}
and discussed in Section~\ref{sec:subset}.

\section{Infrared properties of X-ray sources\label{sec:ir-prop}}

\begin{figure}
\plotone{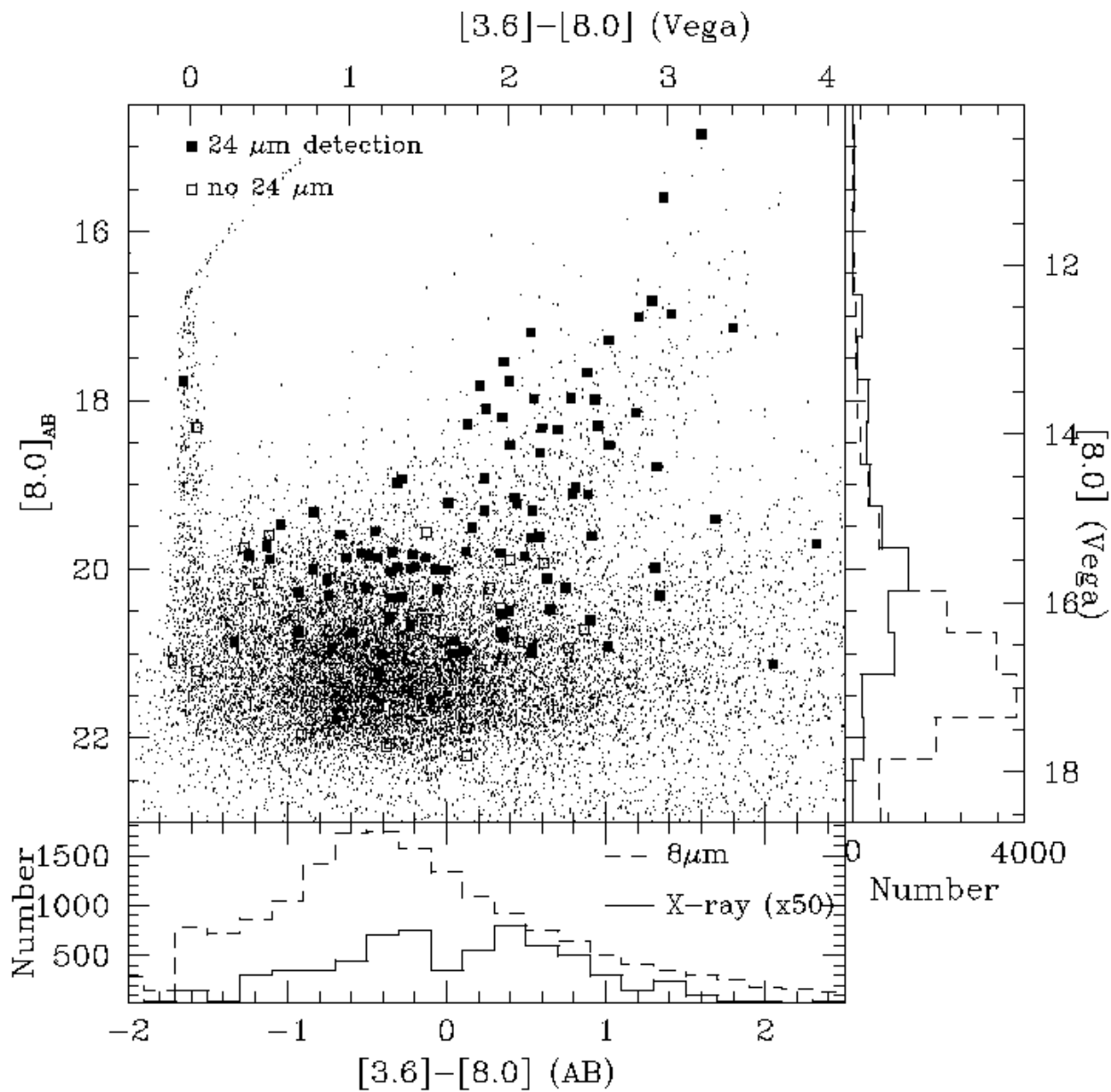}
\caption{IRAC color-magnitude diagram of X--ray sources (squares) and comparison
8\mice-selected EGS sources (dots). Filled squares are sources with 24\mic detections
while open squares are MIPS non-detections. Histograms show color distributions (bottom)
and magnitude distributions (right), with X--ray sources as dashed lines and 8\mic sources 
as solid lines. The X--ray histogram is multiplied by 50 for easier comparison.
\label{col_mag}}
\end{figure}

To understand how the mid-IR properties of X--ray and non-X--ray sources
differ, we compare the properties of the X--ray sources to unbiased samples
selected from the full EGS dataset. There are about 4400 8\mice-selected
sources in the X-ray survey area, so $\sim3$\% of these are X--ray sources. 
In their optical/IRAC/MIPS identification of X--ray sources in the ELAIS-N1 region,
\citet{fra05} found that $\sim12$\% of 8\mic sources were {\it Chandra}-detected,
and that the fraction of X--ray detections decreased with decreasing IR flux.
An IRAC color-magnitude diagram (Figure~\ref{col_mag}) shows a sequence of bright, blue sources with colors
near $([3.6]-[8.0])_{AB}\approx-1.6$, which corresponds to $([3.6]-[8.0])_{\rm Vega}\approx0$. These are 
the colors expected for stars; most of the X--ray and comparison sources are redder
and consistent with galaxy colors \citep[see also Figure~5c of][]{irac_shallow}.
Four X--ray sources (x42, x98, x113, and c152) lie on the blue sequence. x113 is also very blue in
$[8.0]-[24]$, while the other three sources are not detected at 24\mice. Based on these data, 
we identify these four sources as Galactic stars and omit them from the following plots 
and analysis.\footnote{\citet{miy04} also identified x113 as a Galactic star.}  
This contamination rate (4/138 = 3\%) is comparable to that found
by \citet{fra05}.

Figure~\ref{col_mag} shows that
the X--ray sources have different color distributions from the comparison samples.
The X--ray sources' IRAC colors are redder than those of the 8\mic sample in all six band combinations and
their IRAC/MIPS colors ($[8.0]-[24]$) are bluer than those of the 24\mic sample. 
KS tests show that all of these differences are statistically significant. 
The color differences are an indication that the X--ray-producing AGN
affects the SED of the infrared counterpart, as expected.
The X--ray sources are also about a magnitude brighter 
than the typical 8\mice-selected IRAC source and about 0.3~magnitudes brighter
than the typical 24\mic source.
For the X--ray sources, median AB magnitudes in the 4 IRAC bands are
(19.85, 19.95, 20.04, 19.98), while the medians for the full EGS 8\mic sample are
(20.91, 21.03, 21.00, 21.14). The median X--ray source 24\mic flux density is 
182~$\mu$Jy, compared to 146~$\mu$Jy for the 24\mic sample.
While it would be interesting if the brighter infrared fluxes indicated that
X--ray sources preferentially inhabit more massive galaxies, there is a more
mundane explanation: the EGS infrared data are deeper than the X--ray data.
We confirmed this by comparing to the Chandra Deep Field-South, where the
much shallower (470~s) IRAC data at 3.6 (8.0)\mic still detects some 90\% (60\%) of the X--ray
sources despite the much deeper X--ray limit in that field.

Mid-infrared SEDs of galaxies containing AGN are expected to be comprised of several 
components which can have different relative luminosities. 
In nearby galaxies, the IRAC bands contain
the Rayleigh-Jeans tail of the stellar emission and may also include
emission from the interstellar polycyclic aromatic hydrocarbon (PAH) features in the 8\mic band
and hot interstellar dust in the 5.8, and 8\mic bands. At $z>0.25$, the PAH features redshift out
of the IRAC bands and at $z>0.6$ the `1.6\mic bump' from the H$^-$ opacity minimum in
old stars begins to redshift through the IRAC bands \citep{se99}. At $z\sim 2$ the PAH
features are redshifted into the MIPS 24\mic band. AGN emission in the mid-IR is
often phenomenologically described by
a red power-law \citep[$f_{\nu} \propto {\nu}^{\alpha}, \alpha<0$; ][]{elvis94}.%
\footnote{A power law SED does not necessarily
imply a non-thermal origin for the emission.  \citet{ffh05} show that
dust emission from a range of dust types and distributions can produce
the observed SEDs from a variety of AGN.}
The utility of mid-IR power-laws for selection of AGN is further discussed by
\citet{aah05}, who presented a study of 24\mic sources with power-law ($\alpha<-0.5$) SEDs
in the Chandra Deep Field South. Based on the galaxies'
optical-IR SEDs and X--ray and IR properties, they concluded that the majority (including
those not detected in X--rays) harbor an AGN. 

\begin{figure}
\plotone{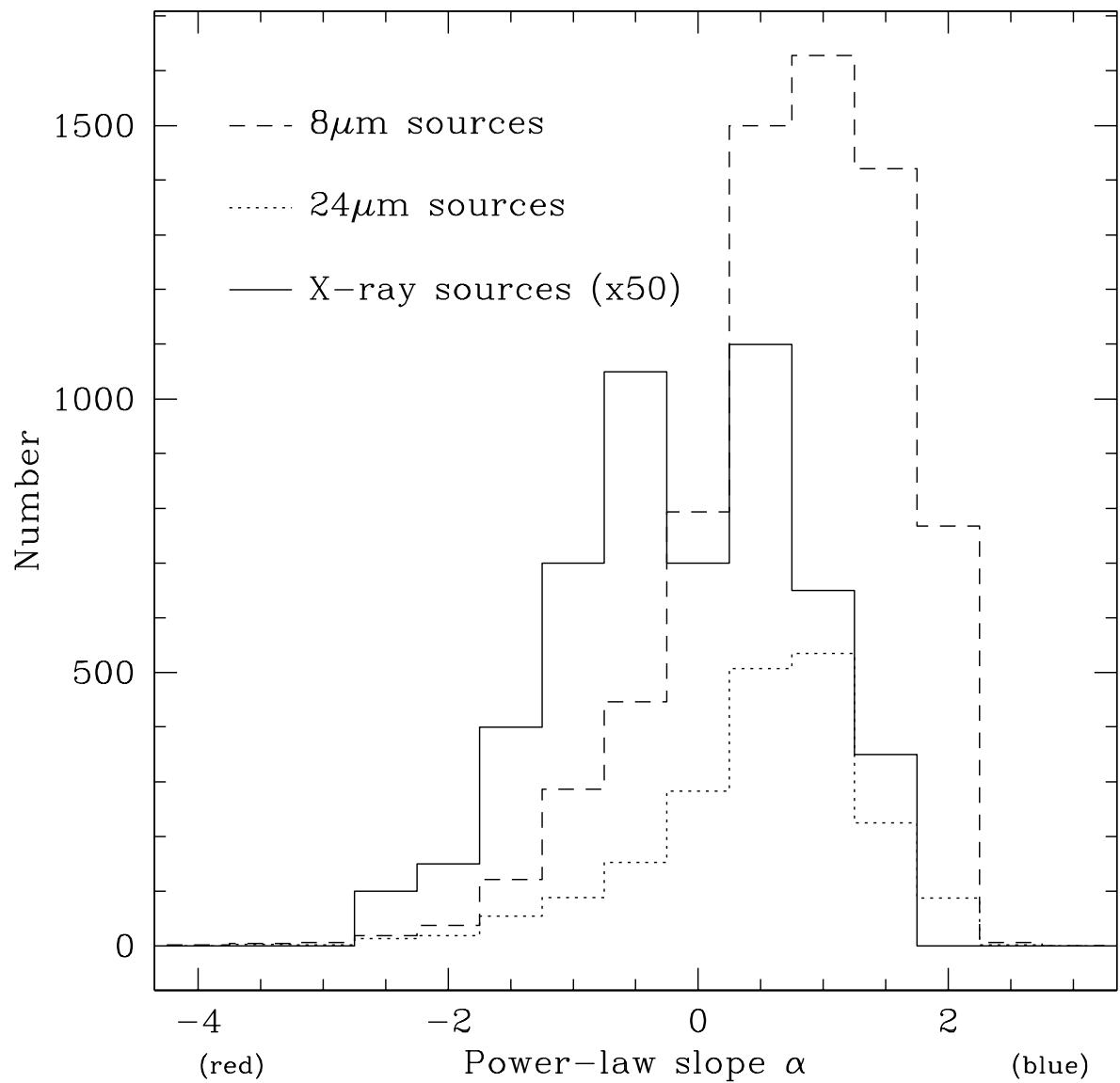}
\caption{Distribution of IRAC power-law spectral indices for X--ray sources
(dashed line), 8\mic sources (solid line), and 24\mic sources (dotted line).
Only objects with acceptable power-law fits are shown: 78\% of the X--ray sample and
$\sim 40$\% of the comparison samples.
The X--ray histogram is multiplied by 50 for easier comparison.
\label{alphas}}
\end{figure}

As a simple description of the EGS X--ray sources' mid-IR SEDs, we attempted to fit
a power law to their IRAC flux densities. We expected that AGN-dominated objects would
have red power-laws with $\alpha<0$, while stellar-dominated 
galaxies would have blue SEDs with $\alpha \approx +2$.
PAH emission at 8\mic or redshifted 1.6\mic bump emission will 
result in poorer fits to a simple power law, as will comparable AGN and
stellar contributions to the total IR luminosity.
Using a simple ${\chi}^2$ method, we fit power laws to the IRAC flux 
densities of the individual sources in the X--ray and comparison samples.
The distributions of power law indices $\alpha$ are shown in Figure~\ref{alphas}.
Seventy-eight percent of the X--ray sources (104/134) had acceptable power-law fits ($P({\chi}^2)>0.01$), 
with about half of these (53) having $\alpha<0$.
(Typical uncertainties in $\alpha$ for the successful fits are $\pm 0.1$.)
Thus {\em only about 40\% of the EGS X--ray sources show the classic red power-law SED in the mid-infrared.}
As might be expected, these sources are fainter than the blue-power-law sources
at 3.6 and 4.5\mic but not different at the longer IR wavelengths.
The fraction of comparison sample sources with acceptable power-law fits is
about 40\% for both the 8- and 24-\mice-selected samples, and their 
median $\alpha$ is much higher ($+0.84$ for 8\mic sources and $+0.64$ for 24\mic sources),
as expected from their bluer colors. 
About 7\% of all 8\mic sources and 9\% of all 24\mic source have good power-law fits with
$\alpha<0$.

AGN and galaxy SEDs are known to be complex, and, as expected, 
power-law SEDs observed in the IRAC bands do not extrapolate well to the 
optical and 24\mic bands. Extensions of the IRAC power-laws to the optical
always over-predicted the $R$-band magnitudes by large factors and, for the
blue sources, under-predicted the 24\mic fluxes by several magnitudes.
In the red sources, the AGN dominates the IRAC bands, so the correlation
might be expected to be better: however, the IRAC power-laws predict the 24\mic values only
to within about 60\%.

The wide range of spectral shapes exhibited by the EGS X--ray sources
means that {\em no proposed mid-infrared color AGN selection will identify all of them.}
Figure~\ref{2colors} compares the IRAC colors of the EGS X--ray sources to those of the 
EGS 8\mic sample in several color-color spaces, where we have marked the regions of IRAC
colors used by various authors to select AGN.
The X--ray sources cover a wide range of colors in all of these plots; 
the X--ray sources not detected at 24\mic tend to have bluer IRAC colors (although this
effect is not statistically significant). 
Figure~\ref{template_2col} shows the predicted colors of several galaxy and AGN templates
which combine optical and near-infrared data from the {\sc hyperz} 
package \citep{bmp00} and mid-infrared data from \citet{lu03}. 
The AGN templates have red colors and lie in the AGN selection regions at all redshifts;
the starburst galaxy templates move into and out of the selection regions depending on $z$;
and the normal galaxy templates are blue at low $z$ but move into the AGN regions
at high redshifts. From this comparison we infer that the blue power-law sources are 
mostly low-redshift galaxies where the galaxy light dominates the mid-IR, while the
red power-law sources are mostly dominated by AGN light but may also include a few starburst galaxies.
As might be expected, the red power-law sources mostly satisfy the various AGN color criteria
while the blue power-law and non-power-law sources mostly do not.

Important properties of any AGN selection criterion are its completeness
(the fraction of all AGN selected) and reliability (the fraction of selected sources which are truly AGN).
For example, \citet{stern05} used optical spectroscopy to determine that some 83\% of the sources 
meeting their mid-IR criterion in the NOAO Deep-Wide Field/IRAC Shallow Survey region
were (mostly broad-lined) AGN. \citet{vg05} matched an X--ray survey of the same
region to the IRAC observations and showed that 67\% of matched X--ray/IRAC sources  
met the \citet{stern05} criterion.
Of course, we cannot directly assess the reliability 
of any color criteria for finding AGN
(because we do not know  how many non-X--ray-detected AGN might be in the survey area),
but we can compute how reliably they select X--ray sources.
The \citet{stern05} criterion seems to be both complete and reliable for shallow
infrared and X--ray data, but it is important to test it against other criteria
and deeper infrared and X--ray data. 

Selecting X--ray sources using published IRAC color criteria does not yield
complete samples. The published IRAC color criteria of
\citet{lacy04}, \citet{stern05}, and \citet{hatz05} select 98, 68, and 38 of 
the 134 EGS non-stellar X--ray sources, respectively, so their completeness is 73\%, 51\%, and 28\%. 
The \citet{hatz05} criterion specifically addressed the selection of type~1 AGN only, and this may
be the cause of its lower completeness.
Detection at 24\mic 
only slightly increases the probability that a source will fulfill the IRAC color criteria.

Reliability is more difficult to assess.
Of all the EGS sources in the survey area (not just the X--ray detections), the color
criteria select 2015, 775, and 580 sources respectively, so their apparent reliabilities are 
5\%, 9\%, and 7\%.  If all of the color-selected sources are truly AGN, 
X--ray undetected AGN would outnumber detected AGN by more than 10 to 1.
This is not completely unreasonable, since we have shown above that our infrared
data are deeper than the X--ray data.  
Also, \citet{donley05} recently found X-ray undetected radio-excess AGN
in the Chandra Deep Field North, 60\% of which are X--ray undetected even in {\it Chandra} exposures $>5$ times 
deeper than the EGS data.  Only about half of the Chandra Deep Field South
AGN selected by the IR power-law criterion of \citet{aah04} are X--ray-detected.

How can we reconcile our results on color selection with the high reliability
and completeness found by \citet{stern05} and \citet{vg05}? Our low reliabilities
in finding X--ray sources are likely due in part to the EGS infrared data being deeper 
than the X--ray data as discussed  above. The much greater depth of the
EGS {\it Spitzer} data compared to the IRAC Shallow Survey data used by \citet{stern05} 
and \citet{vg05} also means that the two surveys probe different redshift regimes. As 
Figure~\ref{template_2col} shows, AGN and normal galaxies have different IRAC
colors at $z\lesssim 2$, but at higher redshifts both types of sources 
have colors meeting the AGN selection criteria.
Also, \citet{stern05} may have found an artificially low contamination rate
because they computed it only from {\em sources with optical spectroscopy}.
Their optical spectroscopy is sensitive to galaxies mostly at $z\lesssim 0.6$, 
meaning that higher-redshift galaxies in the AGN selection area would not have been
included in their sample. Assuming that local templates faithfully represent galaxy and
AGN SEDs at high redshift, mid-infrared color selection of AGN will be more reliable
for samples which do not contain large numbers of high-redshift galaxies.
X--ray, optical, and infrared selection all have different
contributions to make to AGN selection, but no method is both reliable and complete.

\section{Infrared and X--ray properties\label{sec:irx}}

Comparison of X--ray and infrared fluxes is important for understanding the nature of
X--ray sources. X--ray-to-infrared ratios, particularly $f_{\rm 2-10~keV}/f_{24}$,
distinguish between AGN and starburst galaxies; the latter should be bright in
the IR, but very faint in hard X--rays. Figure~\ref{xray_ir}a compares $f_{\rm 2-10~keV}$ 
and $f_{24}$ for the EGS sources: about two-thirds of the X--ray sources ($N=96$) have 
both hard X--ray and 24\mic detections. 
The distribution of $f_X/f_{\rm 24}$ is consistent with AGN powering
the X--ray sources' emission at both wavelengths, as also found by \citet{aah04} 
and \citet{fra05}. The two objects with $f_{24}>5$ mJy
are c72 and x013, discussed in Section~\ref{sec:scuba}.

X--ray hardness ratios are often used to indicate AGN type, with harder
spectra indicating more obscuration. We use $HR=+0.55$ to distinguish
between X--ray-hard and soft sources; this is the equivalent of $HR=-0.2$ for a hardness 
ratio computed from {\it Chandra} counts as used by \citet{szo04} and \citet{rigby04}.%
\footnote{We have not attempted to account for $K$-corrections, which can
be substantial for hardness ratios: see Figure~8 of \citet{szo04}.}
The ratio of soft to hard sources in our sample is about 2:1.
A simple picture of AGN classification predicts that X--ray-hard sources should have 
lower $f_{hX}/f_{24}$ ratios \citep[since the obscuring dust should increase
the 24\mic flux and absorb the hard X--rays;][]{awaki91}. To first order, this
effect should not depend on redshift since flux ratios are distance-independent.
Recent work \citep{lutz04,rigby04} has found very little difference
in $f_{hX}/f_{24}$ between X--ray-hard and soft sources; in the EGS sample,
we also find no statistically significant difference.

The effect of obscuring dust will vary in the IRAC bands:
in the shorter wavelengths, absorption would decrease $f_{\rm IR}$
while at the longer wavelengths, dust emission could increase $f_{\rm IR}$.
We might therefore expect $f_{hX}/f_{\rm IR}$ to be similar for 
obscured and unobscured sources at 3.6\mice, and lower for 
obscured sources at 8\mice. In fact, we find no difference in
$f_{hX}/f_{\rm IR}$ in the IRAC bands between X--ray-hard and soft sources.
[$f_{hX}/f_{3.6}$ is higher for IR-red sources compared to blue sources,
but this merely reflects the fact that red sources are fainter at 3.6\mic (Section~\ref{sec:ir-prop}).]
Comparing the full-band X--ray fluxes (rather than only the hard band) to the IRAC
flux densities (Figures~\ref{xray_ir}b and \ref{xray_ir}c), there {\em is} a statistically
significant difference in $f_X/f_{\rm IR}$ between X--ray-hard and soft sources:
the hard sources have lower $f_X/f_{\rm IR}$ at both 3.6 and 8.0\mice.
This is not quite consistent with the expectations given above, or with
the results from 24\mice, an indication that diagnostics such as $f_X/f_{\rm IR}$ 
likely oversimplify the complex physical processes responsible for
IR and X--ray emission by galaxies and AGN.

Correlations between $f_X$ and $f_{\rm IR}$ arise partly because fluxes indirectly measure distance, 
but nevertheless indicate whether the emission at the two wavelengths is produced
by the same mechanism. Since we have posited that the sources with red IRAC
power laws are dominated by the AGN, we expect that they will show better 
correlation of $f_X$ (which is also AGN-dominated as shown earlier) with 
$f_{\rm IR}$ than the non-red power-law sources.
Figures~\ref{xray_ir}b and \ref{xray_ir}c show that, as expected, the correlations
of $f_X$ with $f_{\rm IR}$ are poor for the blue power-law sources and stronger
for the red sources and the full sample. The correlation strength increases slightly with 
IRAC band wavelength \citep[a similar effect was found by][]{fra05}.
Including the many IRAC-red power-law sources undetected in X--rays \citep[also cf.][]{aah05} 
in comparisons of $f_X$ and $f_{\rm IR}$ would worsen the correlation. 

The IRAC power-law indices can be used to define two
different object types: the red sources, dominated by the AGN light,
and the blue sources, dominated by galaxy emission. The difference
between AGN- and galaxy-domination of the SEDs could be due to 
differing amounts of AGN obscuration, a difference in the relative 
intrinsic luminosities of the two components, or some combination of these.
If the blue sources are blue because obscuration causes absorption
of the AGN's IR light, we would expect them to have harder X--ray spectra 
and lower $f_X/f_{\rm IR}$ ratios. If the blue sources are IR-blue because
their AGN are intrinsically IR-faint, then their X--ray hardness
should be the same as the red sources' and their $f_X/f_{\rm IR}$ should
be the same or higher. Figures~\ref{xray_ir}b and \ref{xray_ir}c
show that the blue sources do have higher $f_X/f_{\rm IR}$, but only 
at 3.6\mic (where the red sources are fainter by definition).

A direct comparison of X--ray hardness
and mid-IR power-law index (Figure~\ref{hr_alpha}) shows no correlation between hardness ratio and
$\alpha$. X--ray soft sources with power-law SEDs are more likely to be red
(the median $\alpha$ for soft sources is $-0.30$ and that for hard sources is $+0.15$),
but the $\alpha$ distributions of soft and hard sources are not significantly different.
The distributions of hardness ratios of the red and blue power-law sources and
the sources with poor power-law fits are also not significantly different.
Fitted column densities $N_H$ are available for about half the sample 
from \citet{geo05}, but using these 
instead of hardness ratios does not change these results. 
While there is a connection between X--ray hardness ratios and mid-infrared spectral indices,
the latter do not appear to be particularly good indicators of AGN type.
Rather, they measure the degree to which the mid-infrared is dominated
by AGN emission and correlated with X--ray flux.

\section{Properties of X--ray cross-identifications\label{sec:subset}}

\begin{figure}
\plotone{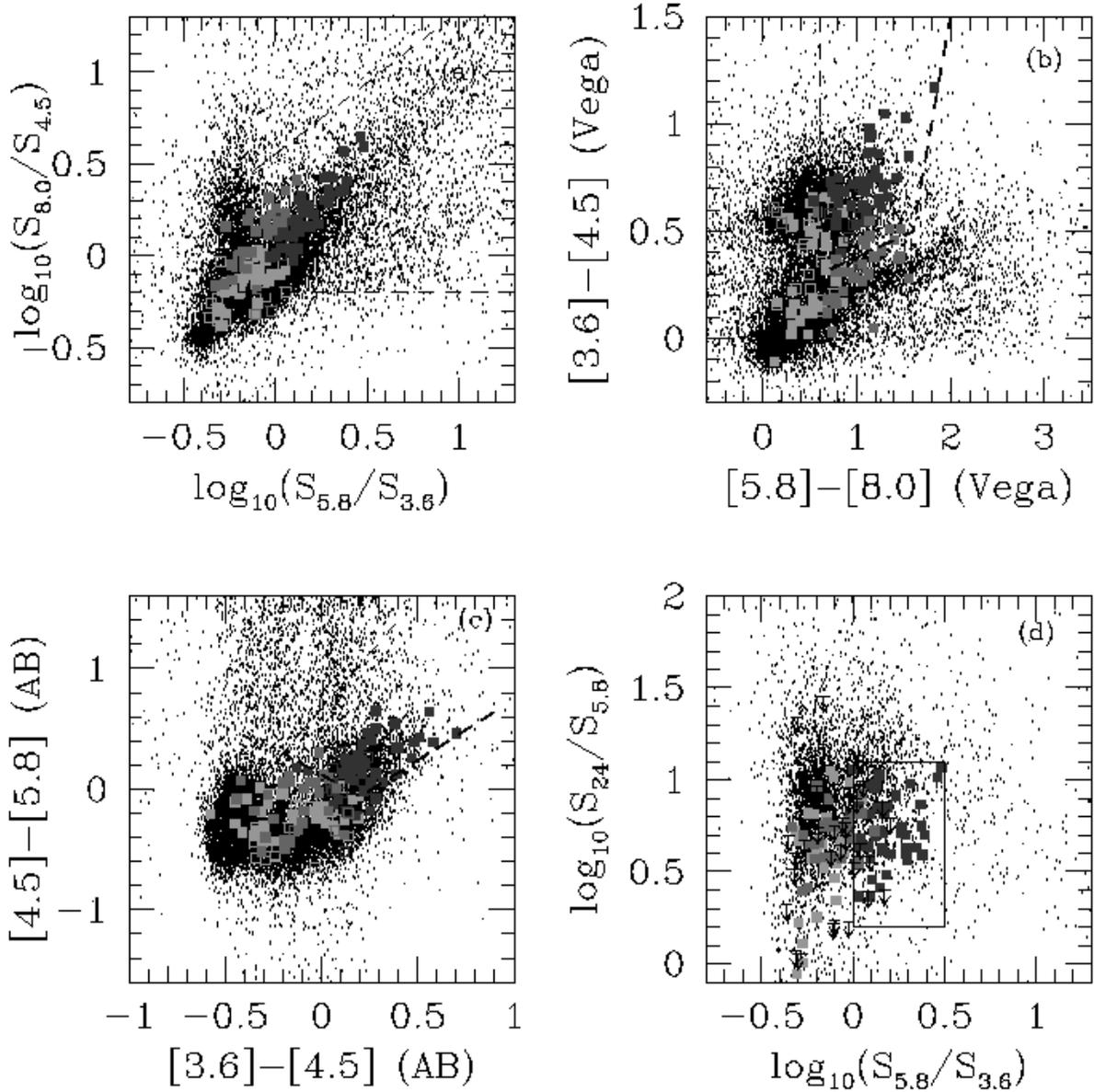}
\caption{Colors of comparison EGS sources (dots) and X--ray sources (squares). 
Filled (open) squares are X--ray sources detected (undetected) at 24\mice.
Darkest shading is for successful power-law fits with $\alpha<0$; lightest shading for
successful power-law fits with $\alpha>0$; intermediate shading for unsuccessful power-law fits.
The AGN selection regions are surrounded by dashed lines.
(a) IRAC colors used by \citet{lacy04}; (b) IRAC colors used by \citet{stern05};
(c) IRAC colors used by \citet{hatz05}; (d) IRAC$+$MIPS~24\mic colors used by \citet{lacy04}; box indicates
approximate location of SDSS quasars and obscured AGN candidates in that paper.
\label{2colors}}
\end{figure}

\begin{figure}
\plotone{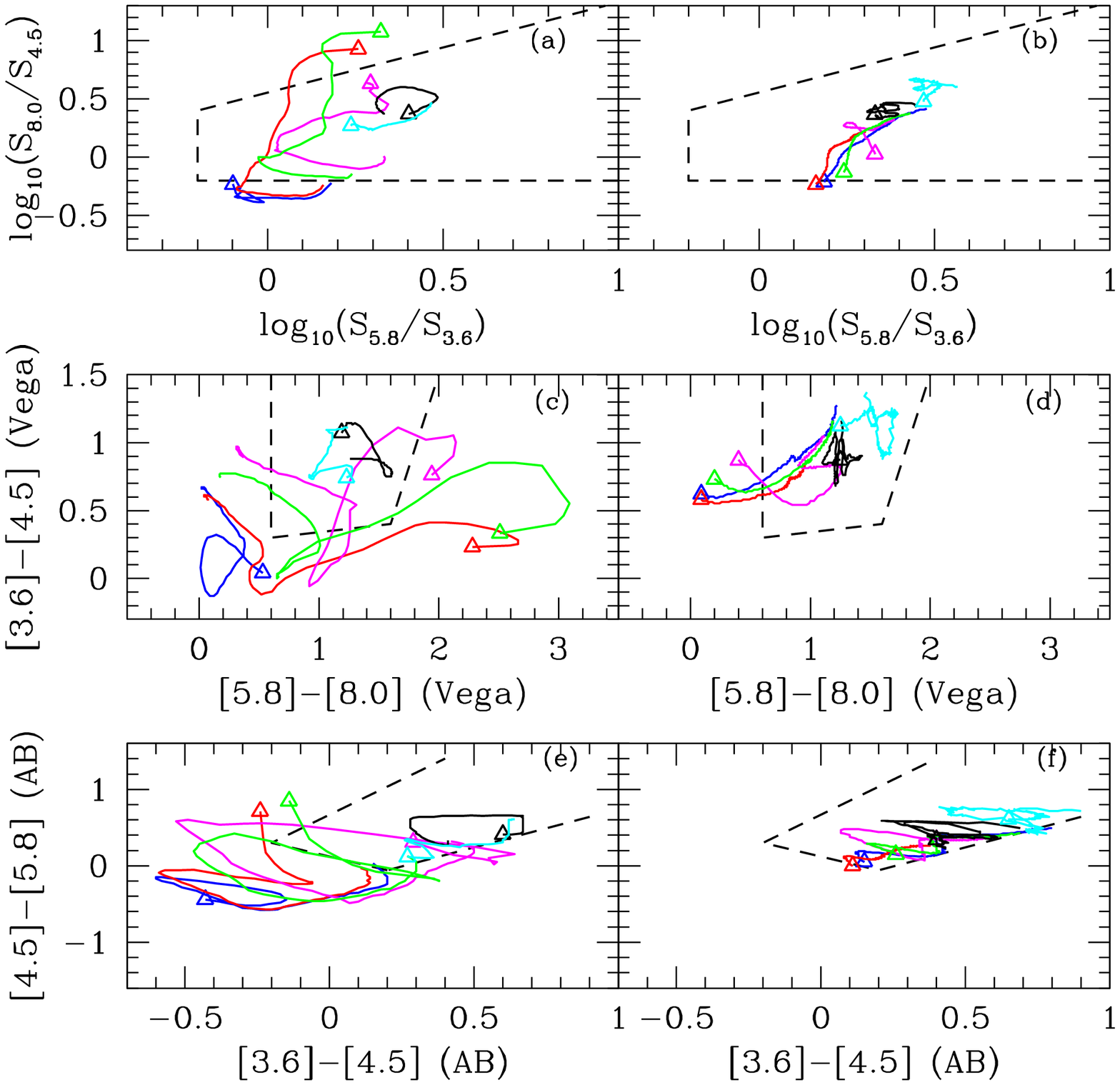}
\caption{Predicted IRAC colors for different galaxy and AGN templates.
Left column is $z=0$ to $z=2$, right column is $z=2$ to $z=7$. 
Open triangles mark the lowest redshift on each track. 
The templates combine optical and near-infrared data from the {\sc hyperz} 
package \citep{bmp00} with mid-infrared data from \citet{lu03}. 
Blue: \citet{cww} elliptical, red: \citet{cww} Scd, magenta: Arp~220, green: M82,
black: the prototype AGN NGC~1068, cyan: the Seyfert galaxy NGC~5506.
Panels (a) and (b) show IRAC colors and AGN selection regions as in Figure~\ref{2colors}a;
panels (c) and (d) as in Figure~\ref{2colors}b; panels (e) and (f) as in Figure~\ref{2colors}c.
\label{template_2col}}
\end{figure}

\begin{figure}
\plotone{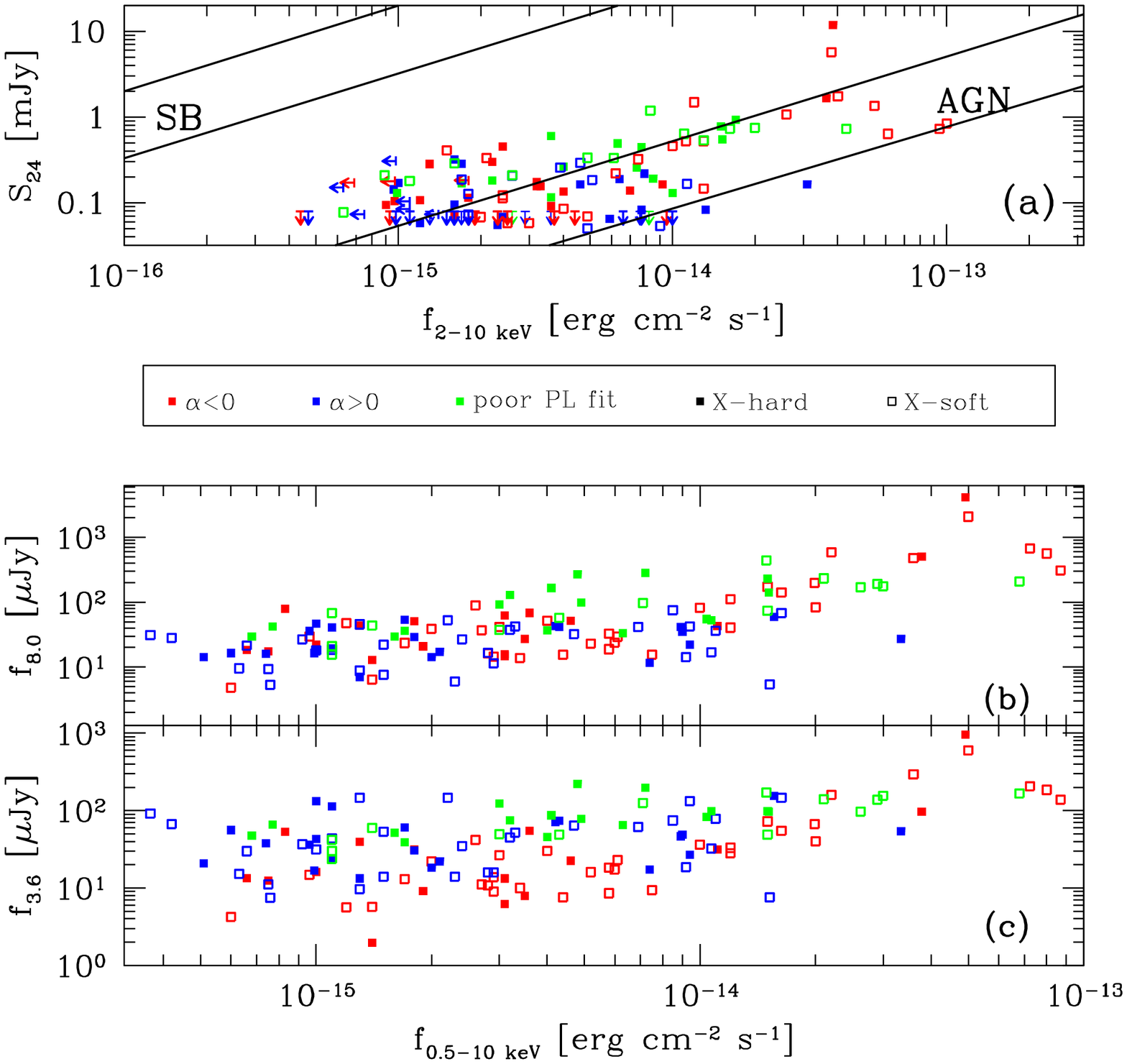}
\caption{(a) Hard X--ray vs. 24\mic flux densities for EGS sources, adapted from Figure~1
of \citet{aah04}. `AGN' region is extrapolation of local hard-X--ray-selected AGN \citep{pic82};
`SB' region is extrapolation of local starburst galaxies \citep{rcs03}.
(b) Full-band X--ray flux vs. IRAC 8.0\mic flux density.
(c) Full-band X--ray flux vs. IRAC 3.6\mic flux density.
Red symbols are successful IRAC power-law fits with $\alpha<0$; blue symbols
are successful power-law fits with $\alpha>0$; green symbols are unsuccessful power-law fits.
Filled symbols are X--ray-hard sources ($HR>+0.55$), open symbols are X--ray-soft sources.
\label{xray_ir}
}
\end{figure}

\begin{figure}
\plotone{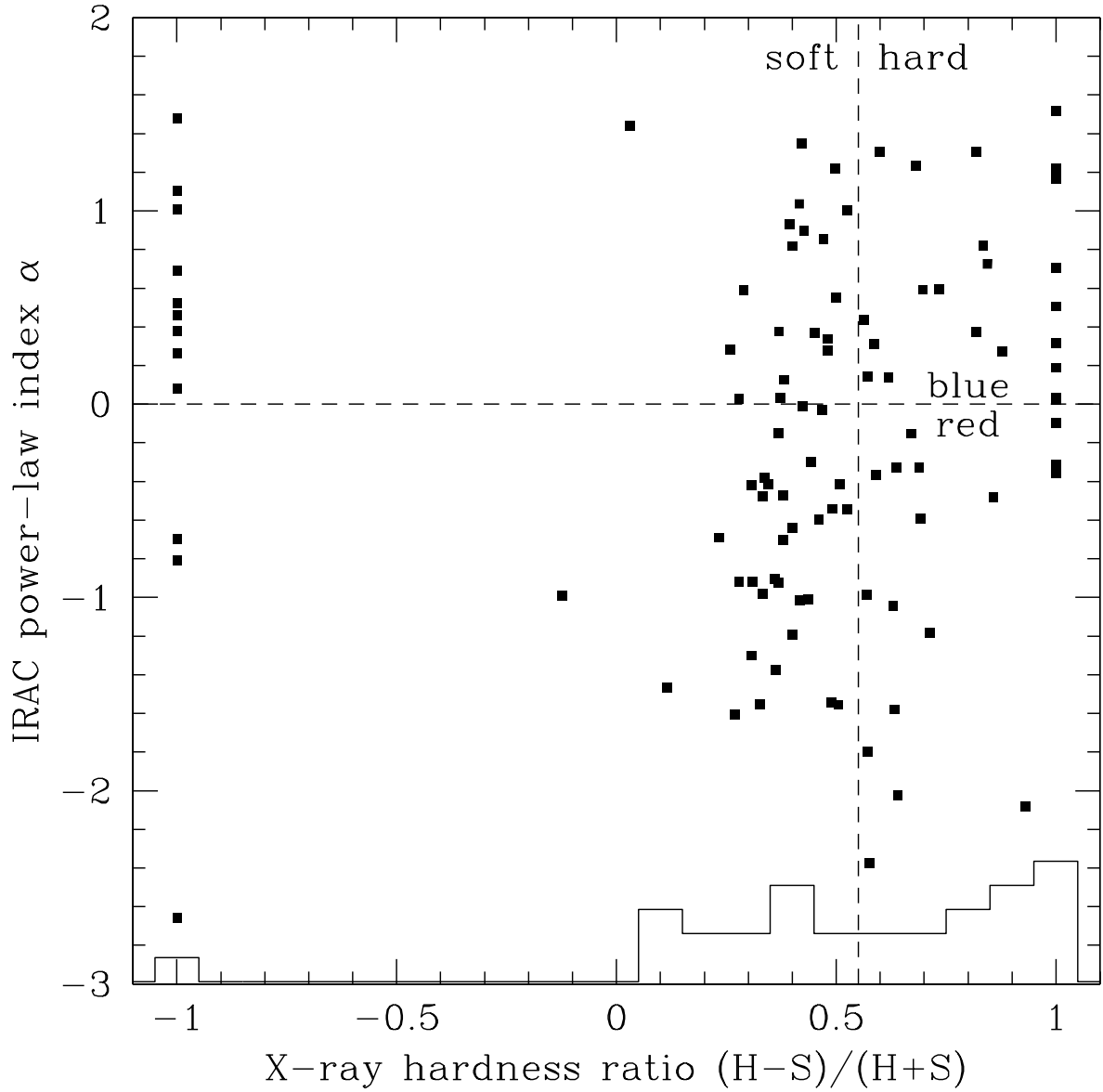}
\caption{X--ray hardness ratio vs.\ mid-IR power-law index $\alpha$. Histogram at bottom
shows distribution of HR for X-ray sources with poor mid-IR power-law fits.
\label{hr_alpha}}
\end{figure}

\begin{figure}
\plotone{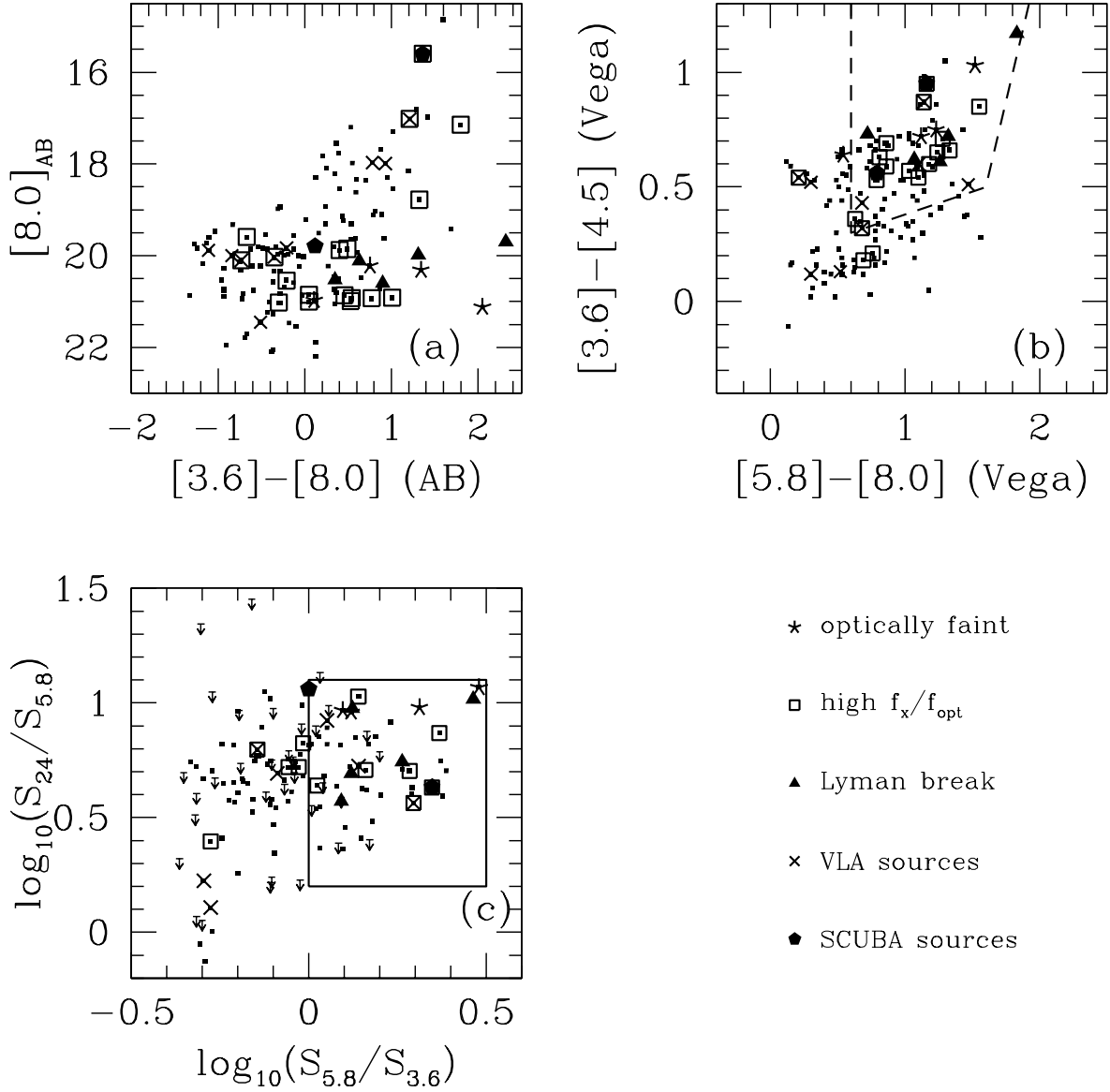}
\caption{IRAC/MIPS color-magnitude diagrams of subsets of X--ray sources. 
Stars: optically faint sources; boxes: high \xopt sources; triangles: Lyman-break sources;
crosses: VLA sources; pentagons: SCUBA sources. Small black squares: remaining X--ray sources.
\label{cmd_special}}
\end{figure}

Many of the EGS X--ray sources are detected at other wavelengths as part of
other surveys. 
Figure~\ref{cmd_special} shows versions of Figures~\ref{col_mag}, \ref{2colors}b,
and \ref{2colors}d where the different classes of sources,
including Lyman-break sources \citep{ccs03}, radio sources \citep{egs_vla91}, 
SCUBA sources \citep{egs_scuba,webb03}, and high \xopt objects
are explicitly marked. They clearly have a wide range of infrared properties, which
we examine in more detail below. We caution that the lower spatial resolution of 
the {\it Spitzer} imaging compared to, e.g.,
optical data means that source confusion could be an issue when IR properties
are derived.

\subsection{Lyman-break sources}

A catalog of Lyman-break sources (LBS) in the EGS is given by \citet{ccs03};
it includes 253 objects in the area covered by the IRAC imaging. Most of the
spectroscopically-identified sources are classified by \citet{ccs03} as galaxies, but there are also three
(narrow-lined) `AGN' and three `QSOs' (broad-lined AGN). \citet{egs_lbg}
find that the LBS have a wide range of properties, with about 5\% being
`infrared-luminous', dusty star-forming galaxies. The QSO/AGN-classified LBS
are among the brightest in the IRAC bands, and tend to be bluer in both
$R-[3.6]$ and $[8.0]-[24]$, indicating that their infrared luminosities
are not due to star formation.
\cite{nls05} identified matches between EGS Lyman-break and X--ray sources
and used these to estimate the evolution in the space density of moderate-luminosity 
AGN from $z=0.5$ to $z=3$.
Here we discuss the infrared properties of the five matches between
sources listed in the \citet{egs_cxo} and \citet{ccs03} catalogs.

The LBSs detected in the mid-infrared and X--rays include
all three of the \citet{ccs03} QSOs (sources c82, c99, c113),
one AGN (c104), and one galaxy (c128).
The two remaining \citet{ccs03} Lyman-break AGN are not detected in 
X--rays or at 24\mic and are also infrared-fainter than the X--ray sources, 
with $[3.6]\approx24$.
The five X--ray detected sources have X--ray fluxes ranging from $1-6\times10^{-15}$\cgs,
with hardness ratios $HR\sim0.5$ (except for c113 which is X--ray-faint and
not detected in the hard X--ray band).
Compared to other X--ray sources, and particularly to
sources with similar X--ray fluxes, the LBSs are infrared-faint and red. All have red power-law SEDs.
Source c113 is the IR-reddest X--ray source in the sample, with the steepest IRAC power-law
($\alpha=-2.65$). Source c128, classified as a galaxy, has the bluest IRAC colors of the
LBS but still has a red power-law ($\alpha=-0.36$). All of the LBSs have IRAC
colors consistent with those expected of a galaxy or AGN at high redshift. 
There are about two dozen other X--ray sources with similar X--ray fluxes and colors
and some of these could also be high-redshift AGN.
The infrared observations of the LBG/X-ray sources show that they have low amounts of
both extinction and dust emission.

\subsection{SCUBA and X--ray-bright sources\label{sec:scuba}}

Two SCUBA sources \citep{webb03} are definitively identified with X--ray sources: CUDSS~14.13 is
X--ray source c72, and CUDSS~14.3 is X--ray source c111.\footnote{Two additional SCUBA sources
are identified with infrared/X--ray sources by Ashby et al.\ (2005, submitted). We do not discuss them here
because the Ashby et al.\ infrared/X--ray identifications differ from the optical IDs given by \citet{webb03}, and
are in fact the same as two of the Lyman-break sources (c113, c128) discussed in the previous section.}
These sources differ by about a factor of two in sub-millimeter flux but by almost 
a factor of 40 in X--ray flux.
(c72 is among the brightest X--ray sources, with $f_{\rm 0.5-10~keV} \approx 5\times 10^{-14}$\cgs.)
Both are also radio sources \citep[sources 15V23 and 15V24 in][]{egs_vla91}.
For comparison, we also discuss in this section source x013, which has
comparable X--ray flux to c72 but unknown sub-mm flux.

\begin{figure}
\plotone{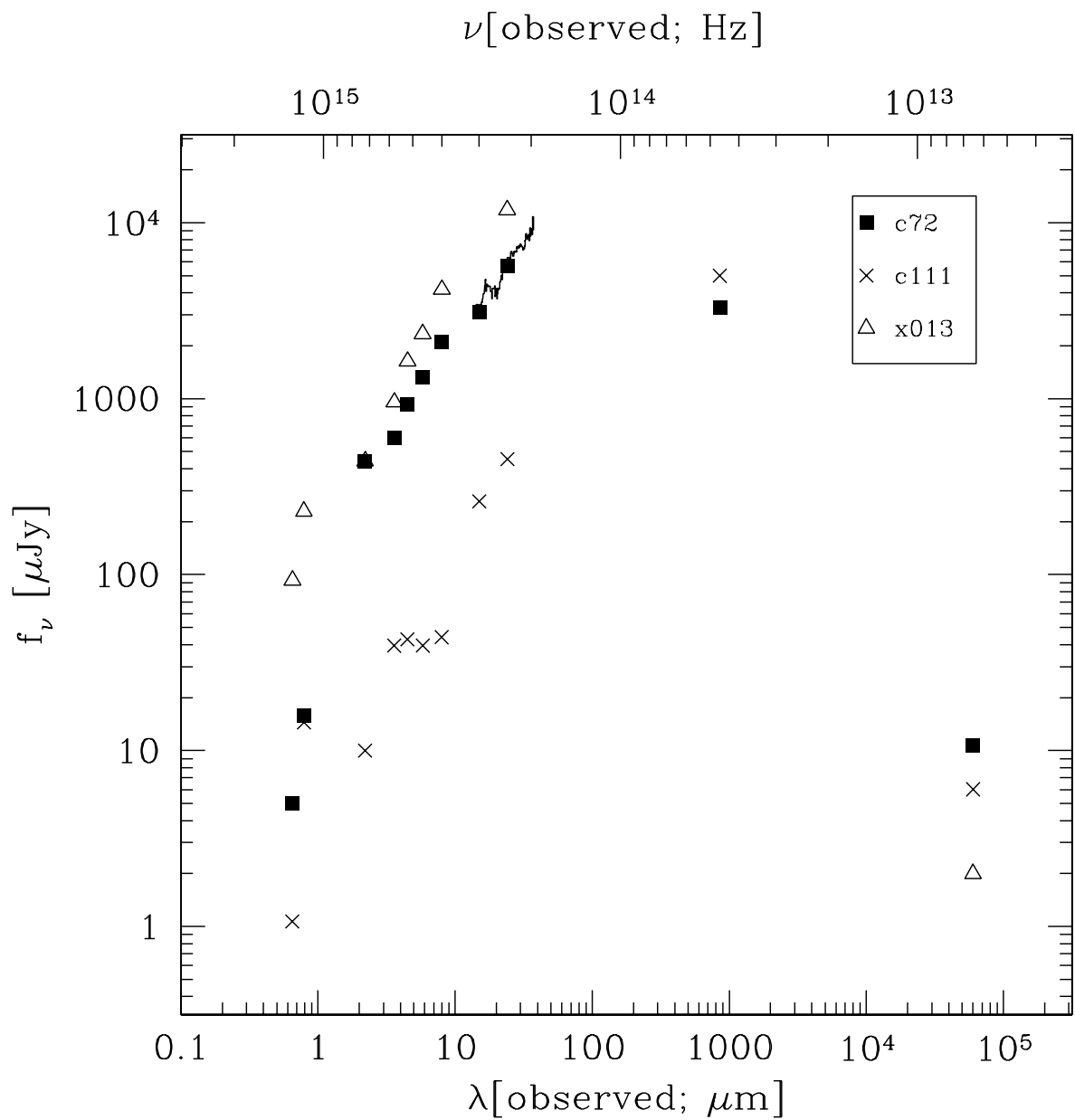}
\caption{Optical and mid-IR spectral energy distributions as a function of observed
frequency for two SCUBA sources and one X--ray-bright source.
Squares: c72 \citep[CFRS~14.1157, shown with IRS spectrum from][]{higdon04}, crosses: c111,
triangles: x013. $I$ and $K$-band fluxes and ISO 15\mic fluxes from \citet{webb03} for c111 and c72;
$I$ and $K$-band from \citet{miy04} for x013.
\label{xrb_sed}}
\end{figure}

The three sources c72, x013, and c111 provide an interesting set of contrasts.
Figure~\ref{xrb_sed} shows the three sources' optical-to-radio spectral energy distributions.
Although both are SCUBA sources, c72 and c111 have few other properties in common; c72 is
much more similar to the non-SCUBA source x013. The latter two sources are both
X--ray and infrared-bright --- they are the two bright, red sources in the
upper right of Figure~\ref{col_mag} --- and have much higher ratios of IR-to-X--ray 
flux density than the other EGS sources (see Section~\ref{sec:irx}). 
Both have red IRAC power-law SEDs, and fall into the `AGN boxes' in all of Figures~\ref{2colors}a--\ref{2colors}c.
Their X--ray hardness ratios are not particularly extreme (0.49 for c72, 0.57 for x013; 
see Figure~\ref{hr_alpha}). While the mid-IR flux densities of c72 and x013 are quite similar,
they differ in the $R$ band by more than 3 magnitudes. This could be due to different
amounts of extinction or simply different redshifts (the $R$-band samples the rest-UV for c72).
By comparison, c111 is much fainter in both the infrared and
X--rays ($[3.6]=19.9, f_{\rm 0.5-10~keV}=1.3\times10^{-15}$\cgs), undetected in the soft X--ray band,
has a nearly flat SED in the IRAC bands ($\alpha=-0.1$) and a very red $[8.0]-[24]$ color. 
These properties suggest that c111 may have significant obscuration.

Source c72 has been the subject of several other studies. 
It was included in the Canada-France Redshift Survey as CFRS~14.1157
\citep[spectroscopic redshift $z=1.15$; ][]{cfrs_ham}. 
Using this redshift and ${\Omega}_m=0.3, H_0=70$~km~s$^{-1}$~Mpc$^{-1}$, the X--ray
luminosity is $L_X = 4\times10^{44}$~erg~s$^{-1}$, making this source a `QSO' as defined by
\citet{szo04}. \citet{higdon04} presented the mid-infrared spectrum of c72: 
the spectrum is essentially a power-law with $\alpha \sim-1.1$ (somewhat shallower
than the IRAC index, $\alpha=-1.5$). The lack of PAH features in the spectrum led \citet{higdon04}
to conclude that this source is AGN-dominated in mid-IR wavelengths.
Their estimate of the total IR luminosity ($10^{13}L_{\sun}$) would classify it as 
a hyper-luminous infrared galaxy.
\citet{was03} concluded that the sub-millimeter flux of c72 was too large
to be produced by the AGN and must be produced by star formation.
\citet{webb03} state that HST imaging 
of c72 ``shows an extended object ($\sim$2\farcs5) with multiple components separated by
diffuse emission''.

Less is known about the other two sources discussed here.
\citet{webb03} state that the optical counterpart of source c111 has ``disturbed''
HST morphology and estimate its redshift to be in the range $1.6<z<3.2$.
Source x013 is outside the region covered by most of
the other surveys, including the {\it Chandra} field of view and the SCUBA map.
\citet{was04} describe x013 as having a `stellar' optical profile, but give 
an $I$-band magnitude brighter than their optical image's saturation limit; it is also
near saturation on the Subaru image. \citet{miy04} also describe this source as having
`pointlike' optical morphology and state that its \xopt is consistent with that of an AGN.

These three sources present several puzzles.
What can be producing the `extra' mid-infrared flux compared to the X--rays in c72 and x013, 
particularly in the IRAC bands? Perhaps these sources show the `5 micron excess' noted by
\citet{em86} and attributed by \citet{mr88} to hot dust. 
Another possible explanation is that perhaps the infrared and X--ray fluxes for these two sources
are actually detections of multiple unresolved sources. One source could be an AGN, the true 
X--ray counterpart, while the other (a star-forming galaxy) provides the extra infrared flux.
(The multiple optical components of c72 discussed by \citet{webb03} 
are only about 0\farcs8 apart and would not be resolved in the {\it Spitzer} imaging:
however, they would also not be resolved in the mid-IR spectroscopy, which
showed few signs of star formation.)
Why are c72 and c111, both sub-mm galaxies, so different in the IR and X--rays? 
The multiple-source hypothesis might account for part of the difference between c72 and c111 if the
latter is truly a single source. The difference in IR and X--rays might simply be a result of
different AGN properties, unrelated to the star formation probed by the sub-mm. Also, if the redshift
estimate of \citet{webb03} is correct, c111 is at significantly higher redshift than c72;
the well-known `negative $K$-correction' in the observed sub-mm could account for the similar
sub-mm fluxes but much different IR and X--ray fluxes.
To summarize, mid-IR and sub-mm-bright sources show a variety of X--ray properties.

\subsection{Radio sources}

The eleven X--ray- and IR-detected radio sources in the survey region have a wide range of
properties. They include the two SCUBA sources discussed above, seven sources
detected in the deep VLA survey by \citet{egs_vla91} and one source (x005) outside the
\citet{egs_vla91} area but detected in the shallower  VLA survey by 
Willner et al.\ (2006, in preparation). The radio sources can be divided into
two groups by 8\mic flux density: c72, x005, c61 and c77 have $[8.0]_{\rm AB}<18$ and
red power-law SEDs. The other seven sources have $[8.0]_{\rm AB}\gtrsim 20$ and all but one
have blue power-law SEDs (c111 has a nearly flat SED with $\alpha=-0.1$).
These two groups are similar to those found by Willner et al.\ %
in their study of IRAC counterparts of VLA sources, although the majority of those
sources did not have X--ray observations.
The radio sources cover a wide range in X--ray flux: source c61 is the X--ray-brightest 
in the sample ($f_{\rm 0.5-10~keV}=1.6\times10^{-13}$~\cgs), and source c106 is one of X--ray-faintest. 
The four 8\mice-bright sources are also the brightest in X--rays, although
the X--ray fluxes do not group as neatly as the 8\mic flux densities. There is no clear 
distinction between the two radio source groups in X--ray hardness ratio, nor do the radio/X--ray
sources as a whole appear to be preferentially harder or softer (or have a different $N_H$ distribution)
than the typical X--ray source. By contrast, \citet{geo03} found that radio sources in the Phoenix Deep Survey 
were more X--ray obscured: their survey had almost five times as many objects so perhaps
small numbers are biasing our results.
The radio-detected X--ray sources do not clearly
differ in their infrared or X--ray properties from the non-radio-detected sources.

The radio sources also have a wide range in radio properties, with 5~GHz flux densities
varying by a factor of about 1000 for the \citet{egs_vla91} sources. (Source x005 is 
much radio-brighter, with $S({\rm 5~GHz})=23.7$~mJy, and is also bright in the IR.) The radio
flux densities are not correlated with either the IRAC or X--ray fluxes or
colors/hardness ratios, and the two groups seen in 8\mic and X--ray brightness are
not reflected in the radio flux densities or spectral indices. Radio and 24\mic
flux densities are known to correlate for star-forming galaxies and radio-faint
AGN: \citet{donley05} used this to select a sample of `radio-excess AGN', many of which
were undetected in X--rays. Four EGS sources (c55, c83, c103, and c106) have radio-to-infrared ratios 
$q=\log(f_{\nu}{\rm (24\mu m)}/f_{\nu}{\rm (1.4 GHz)})<0$ 
which would qualify them as radio-excess AGN. All four of these sources are 
X--ray and 8\mice-faint; c55 and c83 are radio-bright while c103 and c106
are 24\mice-faint. Without redshift information, we cannot constrain their
radio luminosities, but if similar to the \citet{donley05} sources, these
sources are `radio-intermediate': that is, normal radio galaxies rather
than blazar-like.
The X--ray and radio-to-infrared ratios of the seven non-radio-excess sources are
consistent with X--ray emission being produced by the AGN and radio emission 
produced by star formation \citep{bauer02}.
X--ray/radio detected AGN show a wide range of mid-IR properties.

\subsection{High X--ray-to-optical ratio sources}

X--ray-to-optical flux ratios (X/Os) have been known for some time to 
be useful indicators of the nature of X--ray sources \citep[e.g.][]{mac88}.
The nature of sources with high X/Os cannot be easily explored with
optical spectroscopy since they are often too faint; such sources are
thought to be candidates for high-redshift, heavily obscured AGN \citep{main05, rigby05}.
Figure~\ref{fx_opt} shows the \xopt for all of the 
IR-detected sources: 22 sources have $\log (f_X/f_R)>1$, including
several VLA sources and the SCUBA source c72. 
The X--ray and optical properties of these sources will be examined 
elsewhere; here we examine whether the high \xopt affects their infrared properties.
Figure~\ref{cmd_special} shows that these sources are not completely
distinct from the bulk of the IR-detected X--ray sources. However, 
KS tests show them to be (statistically significantly) fainter and redder
than the lower-X/O sources. It follows that the high-\xopt sources should
have more negative power-law slopes, and they do: the median $\alpha$
is $-0.54$ for these sources, compared to $+0.14$ for the low-\xopt sources.
\citet{rigby05} studied a comparably-sized sample of optically-faint X--ray sources
in the Chandra Deep Field South and found similar mid-IR properties.

\begin{figure}
\plotone{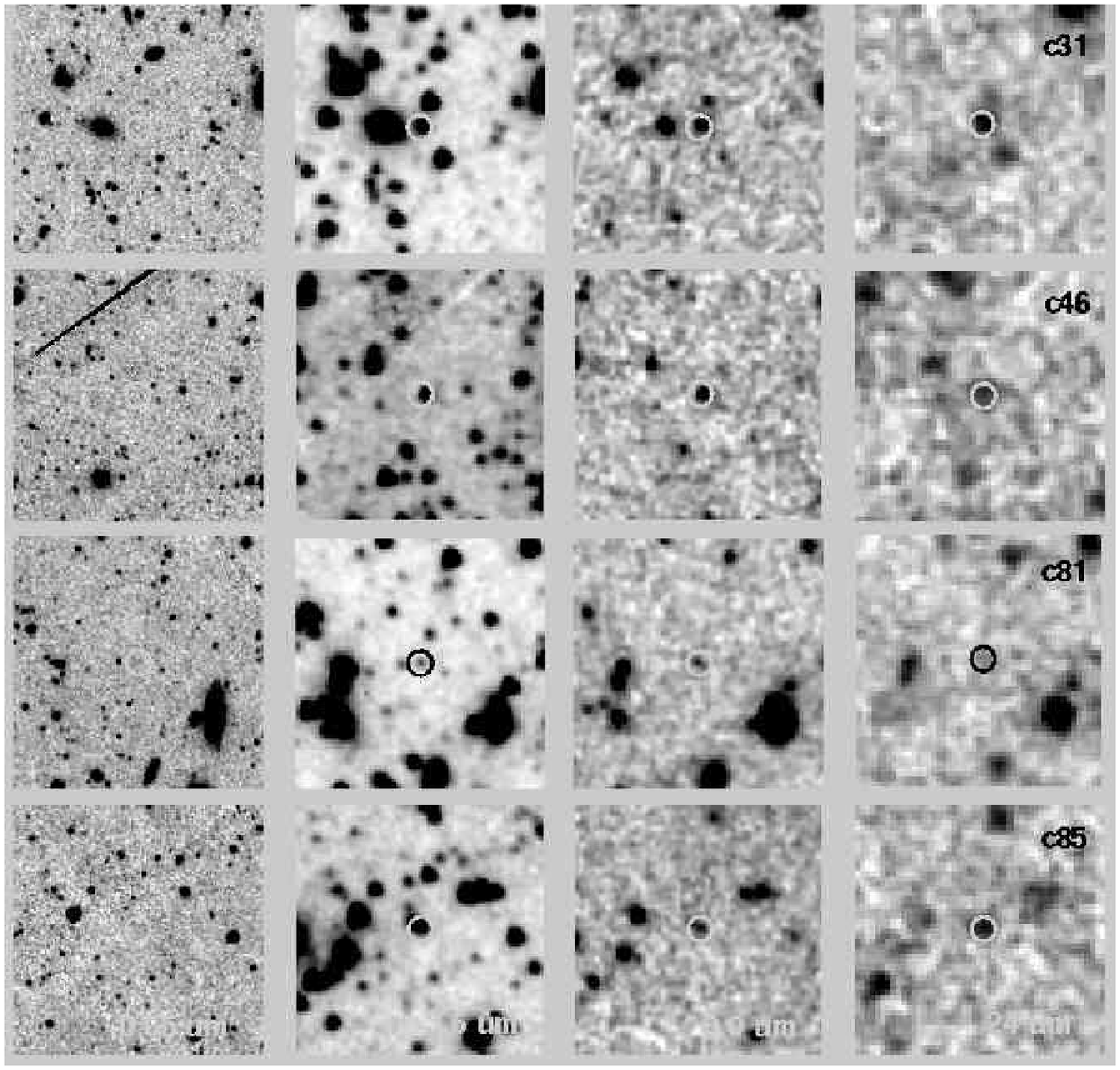}
\caption{Optical ($R$-band) and mid-IR (3.6, 8.0, 24\mice) negative images of the optically-faint 
X--ray sources, centered on the X--ray position. Source names are given on the 24\mic panels.
All images are $1\arcmin \times 1\arcmin$, with North up and East to the left;
circles are 3\arcsec\ in radius.
\label{pix_noopt}}
\end{figure}

Is there anything special about the four IR-detected X--ray sources with 
no optical counterparts? These sources are X--ray faint, with $f_{\rm 0.5-10~keV}<3.5\times 10^{15}$\cgs, 
but should still be reasonably robust X--ray detections ($\gtrsim 10$ full-band net counts).
Source c31 is not detected in the hard X-ray band and the hardness ratios of the 
remaining three sources are about $+0.6$, not atypical. Infrared images of these four sources
are shown in Figure~\ref{pix_noopt}: all are detected at 24\mice, although
only weakly in the case of source c81. The optically-undetected sources are again fainter and 
redder in the mid-IR than the typical X--ray source. Source c81 in particular is 3.5~mag fainter 
at 3.6\mic than the median EGS X--ray source, and accordingly has one of the steepest mid-IR spectral 
slopes of any of the X--ray sources ($\alpha=-2.37$). All four optically-undetected
sources have red power-law IRAC SEDs. Although their $R-[3.6]$ colors make them
`Extremely Red Objects' by the definition of \citet{irac_ero}, such
objects are not unusual. About a third of the X--ray sources and
$\sim60$\% of 8\mic sources (in a shallower IRAC survey) are also EROs \citep{irac_ero}.
Our optically-undetected sources are not directly comparable to the `Extreme X--ray/Optical sources' 
found by \citet{ak04}, since \citeauthor{ak04}'s EXOs are undetected even in the HST $z_{850}$ band, 
($\log(f_X/f_z)\gtrsim2$) while we do not have observations at the longest optical wavelengths.
The EGS optically-undetected sources appear to be merely extreme examples of
the other high-\xopt sources and do not require very high redshifts or reddening to explain their
IRAC colors \citep[see also][]{rigby05}.

\section{X--ray Sources and the Cosmic Infrared Background}

The fraction of the cosmic IR background originating from AGN is
an important parameter in predicting its overall spectrum. The recent
modeling of \citet{smg04} predicts that AGN emission by itself contributes
little ($<5$\%) to the CIRB, but the combined emission from AGN and their
host galaxies is more significant (10--20\%) in the mid-IR.
Their predictions include the contribution of Compton-thick AGN, 
which are weak X--ray sources at energies $<10$~keV but are predicted
to make a substantial contribution to the IR and $>10$~keV X--ray backgrounds. 
The integrated light from X--ray source counterparts 
(provided that most are AGN, which we have shown above to be the case for
the EGS sources) should therefore provide a lower bound to the true CIRB contribution 
of AGN. Since the {\it Spitzer} data do not have the spatial resolution to 
directly separate AGN and galaxy emission, we compare the integrated
light from the EGS X--ray sources to the `AGN+host' predictions of \citet{smg04}.

\begin{figure}
\plotone{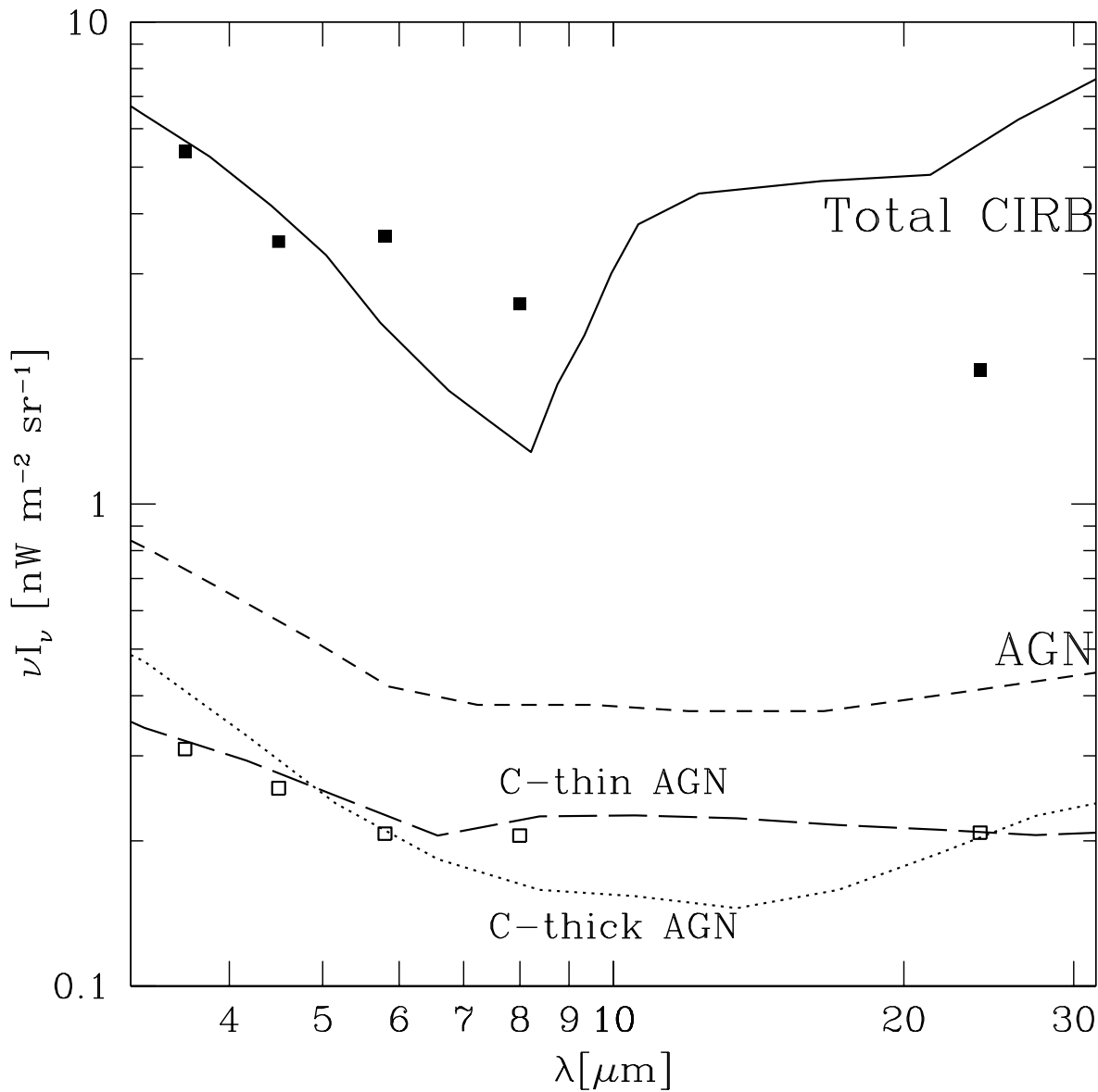}
\caption{Lines: predicted cosmic infrared background from \citet{smg04}. Solid: total CIRB,
short-dashed: CIRB produced by AGN$+$hosts, dotted: CIRB produced by Compton-thick
AGN$+$hosts; long-dashed: CIRB produced by Compton-thin AGN$+$hosts.
Filled squares: measurements of integrated galaxy light \citep{irac_numcts,mips24_numcts};
open squares: integrated light from X--ray sources (this work).
\label{cirb}}
\end{figure}

We computed the integrated mid-IR light from the {\it Chandra} sources alone in
order to have a more uniform X--ray flux limit; the solid angle covered by the
intersection of the {\it Chandra} and {\it Spitzer} images is 1.6$\times 10^{-5}$~sr. 
The integrated fluxes from the EGS {\it Chandra}/{\it Spitzer} sources are 
0.31, 0.26, 0.21, 0.21 and 0.21~nW~m$^{-2}$~sr$^{-1}$ at 3.6, 4.5, 5.8, 8.0, and 24\mice,
with uncertainties of at least 5\% due to the aperture correction, 
absolute flux calibration, and solid angle.
Figure~\ref{cirb} shows the CIRB predictions of \citet{smg04} compared with
our summation of the IR light from the EGS X--ray sources.
We also compare the model predictions to estimates of the integrated galaxy light:
derived by integrating galaxy number counts to limits comparable to the 
EGS data (2--4~$\mu$Jy) in the IRAC bands \citep{irac_numcts} and 60~$\mu$Jy at 24\mic
\citep{mips24_numcts}.
The EGS integrated AGN$+$host fluxes are about half of the \citet{smg04}
predictions, consistent with the notion that these data provide a lower limit
to the true AGN CIRB contribution. The good agreement (within 10\%) of the EGS observations 
with predicted values for Compton-thin AGN$+$hosts suggests, but does not conclusively prove, 
that many of the non-X--ray-detected AGN are Compton-thick.
The agreement between model and observations for the integrated galaxy light is much poorer; in 
particular the \citet{smg04} model under-predicts the CIRB at 5.8 and 8.0\mic and may 
over-predict it at 24\mice.

What is the fractional contribution of AGN to the CIRB? The \citet{smg04} models
predict this to be 7--31\% in the mid-infrared bands with the highest value
at 8.0\mic and the lowest at 24\mice. Dividing the integrated light of the X--ray sources by that of
the galaxy population, we find that the fraction of integrated light from X--ray sources 
is 6--8\% in the IRAC bands and 11\% at 24\mice; there is no
strong change with wavelength. The measured fraction
is about half of the prediction at 3.6 and 4.5\mice, about a third at 5.8 and 8.0\mic
(possibly because of the CIRB under-prediction noted above), and about 1.5 times
the prediction at 24\mic (possibly a fortuitous coincidence, since the models over-predict
the total 24\mic CIRB). 

Our results do not agree with that of \citet{irac_lh}, who found
the fraction of integrated IRAC light from XMM sources in the Lockman Hole 
to increase from 4\% at 3.6\mic to 14\% at 8\mice. 
While those results are based on observations of comparable depth to those
in the EGS, the area covered is small (22 arcmin$^2$) and contains only 
nine X--ray sources, so we do not consider the disagreement worrisome.
\citet{fra05} did not give the integrated flux from their ELAIS-N1
sources in the IRAC bands. To their limiting 24\mic flux of 200 $\mu$Jy, 
\citet{fra05} find that the X--ray sources produce $\sim 10$\% of the 24\mic light,
which they find to be consistent with the $15\pm5$\% at 15\mic found by \citet{fadda02}
and is also consistent with our results.
\citet{brand05} find a similar 10\% fraction to a 24\mic limit of 300~$\mu$Jy.
Although \citet{brand05} find that the fraction of IR light contributed by
AGN decreases as the IR flux limit decreases, and this makes sense given
that AGN should increase the IR flux of a galaxy, there does not seem to be a strong 
evolution in this fraction over the 80--300~$\mu$Jy range in $f_{24}$.
A larger sample of X--ray sources, which will be available from 
the {\it Chandra} observations of the full-length EGS, will enable 
a better determination of the change in AGN fraction with IR flux limit.

AGN and their hosts make a disproportionate contribution to the CIRB in the IRAC bands,
since they make up $<4$\% of galaxies in the sample but contribute 6--8\% of the integrated light. 
At 24\mice, the AGN host galaxies are about 10\% of the total number and contribute 
around 10\% of the integrated light. The wavelength-independence of the integrated
light fraction suggests that X--ray surveys select sources typical of the CIRB contributors
regardless of the source of infrared emission. This is paradoxical given the clear differences
between the SEDs of X--ray and non-X--ray sources, but differences in the redshift
distributions of the two groups might explain the puzzle. Further work on the CIRB
contributions of AGN would be greatly aided by some method of detecting 
Compton-thick AGN among the infrared sources; confirming that X--ray-undetected
IR power-law galaxies are AGN is one possible step forward.

\section{Discussion and Summary}

Of about 150 X--ray sources within the Extended Groth Strip,
more than 90\%  have IRAC counterparts at flux densities $>1-6 \mu$Jy, 
and about two-thirds are detected with MIPS at flux densities $>83\mu$Jy. 
At the flux limits of the X--ray surveys, most of the sources
are expected to be AGN. The ratios of X--ray to optical and IR flux are
consistent with this expectation. 

The infrared SEDs of the X--ray sources 
show a broad range of properties but reasonable agreement with predicted
colors from nearby template objects.
About 40\% of the X--ray sources have a red power-law SED dominated by the AGN. 
The remaining
60\% of the sources have either blue or non-power-law IRAC SEDs, indicating
domination by galaxy light, PAH emission, or a mixture of galaxy and AGN light.
\citet{fra05} found that 39\% of {\it Chandra} sources in the ELAIS-N1 region
had optical/infrared SEDs classified as `QSO' or `Seyfert~1'; if most
of our red sources are also `type 1' (unobscured) AGN, then there is 
good agreement between the two studies.
Published  IRAC color-color criteria select the EGS X--ray sources with 5--9\%
reliability (these values would likely be higher with deeper X--ray observations)
and 25--75\% completeness.

There are good correlations between IR and X--ray fluxes for the
red sources, indicating that the AGN dominates in both wavelength regimes.
We find only marginal evidence (a difference in IR spectral index $\alpha$
between hard and soft sources) for agreement between AGN classifications
based on IR SEDs and X--ray hardness ratios even though the amount of
obscuration should affect the observed properties in both. Variation
in the gas-to-dust ratio and broad ranges of the intrinsic AGN properties
could account for some of the lack of correspondence between IR and X--rays.
Such an explanation is necessary not only for the properties of observed
X--ray sources but to account for the many non-X--ray-detected sources
that have similar IR properties.

The X--ray sources detected at other wavelengths show a wide range of
properties. None are completely distinct from the main X--ray/IR sample,
but most of the subsets are infrared-fainter and redder.
The Lyman-break/X--ray sources are bright and blue when compared to other
LBSs but faint and red (consistent
with being at high redshift) compared to other X--ray sources.
Two sub-millimeter sources are quite different in their
X--ray properties, implying that the properties of the central AGN and 
of the intense star formation producing the sub-mm emission are not necessarily
connected. One of the sub-mm sources and one additional X--ray source are 
`over-luminous' in the mid-IR and we propose several possible explanations.
The radio-detected
X--ray sources divide into two groups based on their mid-IR and X--ray
properties, but these groups do not relate to the radio emission.
The optically faint sources (including 4 sources undetected in the optical)
are again undistinguished in their mid-IR and X--ray properties.

The integrated infrared light from X--ray sources provides a lower bound to the
fraction of the cosmic IR background originating from AGN. Our measurements
of the integrated light from the {\it Chandra} sources are about half of the
predicted AGN$+$host values from \citet{smg04}: as expected, the X--ray
observations do not detect all the AGN.
The amount of integrated light from the EGS X--ray sources is in surprisingly
good agreement with the \citeauthor{smg04} predictions for Compton-thin AGN:
this may be a coincidence (if the X--ray sources contain the right number of
thick and thin AGN) or an indication that most non-X--ray-detected AGN are 
Compton-thick. Disagreements between the observed 
fractions of IR light from X--ray sources and the models at some wavelengths 
may be due to under- or over-predictions of the total CIRB.

Future work on this topic will benefit from the ongoing observations in the
EGS region. Statistics will be improved by the larger, more uniform sample of X--ray sources
soon to become available from the {\it Chandra} observations of the full 2\arcdeg-long EGS.
Spectroscopic redshift information from the DEEP2 survey
will allow determination of luminosities and $K$-corrections such that intrinsic
properties can be computed. Cross-identification with observations at other
wavelengths will allow comprehensive spectral energy distributions to be
constructed. Such SEDs can be compared with population synthesis models to estimate
galaxy stellar masses and star formation histories, which can in turn be compared
to black hole masses computed from X--ray luminosities. 
The large sample of galaxies observed in the EGS no doubt contains many 
non-X--ray-selected AGN, and a better understanding of the X--ray-selected sources
may be of tremendous help in finding these `dark' AGN.

\acknowledgments
We thank the referee for a helpful report and S.~Murray, W.~Forman, M.~Ashby, and 
M.~Pahre for useful discussions.
This work is based on observations made with the {\it Spitzer Space Telescope},
which is operated by the Jet Propulsion Laboratory, California Institute of 
Technology under a contract with NASA. Support for this work was provided by NASA 
through Contract Numbers 1256790 and 960785 issued by JPL/Caltech. 
Based in part on data collected at Subaru Telescope, which is operated by the 
National Astronomical Observatory of Japan. AG and ESL acknowledge support from PPARC.

{\it Facilities:} \facility{Spitzer (IRAC, MIPS)}, \facility{Chandra}, \facility{XMM},
\facility{Subaru}

\clearpage

\begin{deluxetable}{llllrrrrrl}
\tabletypesize{\scriptsize}
\tablecaption{Infrared properties of EGS X--ray sources\label{data_tab}}
\tablewidth{0pt}
\tablehead{
\colhead{CXO\tablenotemark{a}} & \colhead{XMM\tablenotemark{b}} 
& \colhead{IRAC RA} & \colhead{IRAC Dec}
& \colhead{[3.6]} & \colhead{[4.5]} & \colhead{[5.8]} & \colhead{[8.0]} 
& \colhead{[24]} & \colhead{other IDs\tablenotemark{c}}\\
\colhead{} & \colhead{} & \colhead{(J2000)} & \colhead{(J2000)}
& \colhead{AB} & \colhead{AB} & \colhead{AB} & \colhead{AB} & \colhead{AB} & \colhead{}
}
\startdata
\nodata &\object[\[WEG2004\] 14h-142]{142}& 14 15 56.88 & +52 16 07.2 & $19.34\pm0.03$ & $18.87\pm0.02$ & $18.61\pm0.02$ & $18.15\pm0.02$& $17.10\pm 0.03$ &\nodata\\
\nodata &\object[\[WEG2004\] 14h-098]{98} & 14 16 08.45 & +52 21 17.1 & $19.64\pm0.03$ & $20.11\pm0.04$ & $20.57\pm0.07$ & $21.21\pm0.11$& $>19.1        $ &\nodata\\
\nodata &\object[\[WEG2004\] 14h-074]{74} & 14 16 10.59 & +52 16 21.8 & $18.43\pm0.02$ & $18.68\pm0.02$ & $19.11\pm0.03$ & $19.47\pm0.04$& $19.10\pm 0.15$ &\nodata\\
\nodata &\object[\[WEG2004\] 14h-067]{67} & 14 16 13.72 & +52 22 35.6 & $20.15\pm0.04$ & $19.87\pm0.04$ & $19.95\pm0.05$ & $19.81\pm0.05$& $18.36\pm 0.07$ &\nodata\\
\nodata &\object[\[WEG2004\] 14h-132]{132}& 14 16 14.18 & +52 19 39.4 & $20.81\pm0.05$ & $21.11\pm0.06$ & $21.42\pm0.13$ & $21.24\pm0.11$& $19.37\pm 0.15$ &\nodata\\
\nodata &\object[\[WEG2004\] 14h-005]{5}  & 14 16 22.75 & +52 19 16.5 & $18.23\pm0.02$ & $17.83\pm0.01$ & $17.49\pm0.01$ & $17.02\pm0.01$& $16.08\pm 0.02$ &\nodata\\
\nodata &\object[\[WEG2004\] 14h-026]{26} & 14 16 22.94 & +52 12 12.3 & $18.11\pm0.02$ & $17.72\pm0.01$ & $17.38\pm0.01$ & $16.82\pm0.01$& $15.80\pm 0.04$ &\nodata\\
\nodata &\object[\[WEG2004\] 14h-069]{69} & 14 16 29.66 & +52 23 33.1 & $18.92\pm0.02$ & $19.21\pm0.03$ & $19.61\pm0.04$ & $19.59\pm0.04$& $18.62\pm 0.12$ &\nodata\\
\nodata &\object[\[WEG2004\] 14h-091]{91} & 14 16 31.16 & +52 22 10.1 & $20.33\pm0.04$ & $20.59\pm0.05$ & $20.63\pm0.07$ & $20.54\pm0.07$& $>19.1        $ &\nodata\\
\nodata &\object[\[WEG2004\] 14h-032]{32} & 14 16 32.81 & +52 19 01.6 & $18.94\pm0.02$ & $18.56\pm0.02$ & $18.02\pm0.02$ & $17.14\pm0.01$& $15.85\pm 0.02$ &\nodata\\
\nodata &\object[\[WEG2004\] 14h-031]{31} & 14 16 34.53 & +52 22 38.9 & $19.55\pm0.03$ & $19.29\pm0.03$ & $18.89\pm0.03$ & $18.53\pm0.02$& $17.11\pm 0.03$ &\nodata\\
\nodata &\object[\[WEG2004\] 14h-127]{127}& 14 16 34.53 & +52 11 53.5 & $19.57\pm0.03$ & $19.96\pm0.04$ & $20.04\pm0.05$ & $20.31\pm0.08$& $18.36\pm 0.19$ &\nodata\\
\nodata &\object[\[WEG2004\] 14h-070]{70} & 14 16 35.98 & +52 19 11.9 & $19.43\pm0.03$ & $19.55\pm0.03$ & $19.86\pm0.05$ & $19.86\pm0.05$& $18.24\pm 0.10$ &\nodata\\
\nodata &\object[\[WEG2004\] 14h-037]{37} & 14 16 38.08 & +52 23 08.6 & $18.60\pm0.02$ & $18.90\pm0.02$ & $19.33\pm0.03$ & $19.84\pm0.05$& $19.65\pm 0.18$ &\nodata\\
\nodata &\object[\[WEG2004\] 14h-113]{113}& 14 16 39.95 & +52 15 03.9 & $16.12\pm0.01$ & $16.69\pm0.01$ & $17.22\pm0.01$ & $17.77\pm0.02$& $19.03\pm 0.24$ &\nodata\\
\nodata &\object[\[WEG2004\] 14h-042]{42} & 14 16 42.37 & +52 25 26.8 & $19.36\pm0.03$ & $19.87\pm0.04$ & $20.43\pm0.06$ & $21.08\pm0.10$& $>19.1        $ &\nodata\\
\nodata &\object[\[WEG2004\] 14h-013]{13} & 14 16 42.44 & +52 18 12.3 & $16.45\pm0.01$ & $15.87\pm0.01$ & $15.48\pm0.00$ & $14.85\pm0.00$& $13.72\pm 0.01$ &\nodata\\
\nodata &\object[\[WEG2004\] 14h-003]{3}  & 14 16 43.25 & +52 14 35.9 & $17.90\pm0.01$ & $17.79\pm0.01$ & $17.66\pm0.01$ & $17.54\pm0.01$& $16.75\pm 0.07$ &\nodata\\
2       &\object[\[WEG2004\] 14h-063]{63} & 14 16 43.40 & +52 29 02.9 & $21.52\pm0.08$ & $21.34\pm0.08$ & $21.56\pm0.18$ & $20.99\pm0.11$& $19.50\pm 0.20$ &\nodata\\   
\nodata &\object[\[WEG2004\] 14h-146]{146}& 14 16 43.56 & +52 21 02.3 & $18.04\pm0.01$ & $18.23\pm0.02$ & $18.72\pm0.02$ & $17.83\pm0.02$& $16.96\pm 0.04$ &\nodata\\
\nodata &\object[\[WEG2004\] 14h-027]{27} & 14 16 43.97 & +52 22 19.1 & $18.49\pm0.02$ & $18.80\pm0.02$ & $18.99\pm0.03$ & $19.32\pm0.03$& $18.35\pm 0.07$ &\nodata\\
4     &\object[\[WEG2004\] 14h-035]{35} & 14 16 45.39 & +52 29 05.9 & $19.25\pm0.03$ & $18.86\pm0.02$ & $18.74\pm0.02$ & $18.30\pm0.02$& $17.25\pm 0.03$ &\nodata\\
6     &\object[\[WEG2004\] 14h-015]{15} & 14 16 48.78 & +52 25 58.9 & $18.65\pm0.01$ & $18.75\pm0.02$ & $19.04\pm0.02$ & $18.93\pm0.02$& $17.59\pm 0.05$ &\nodata\\
7     &\object[\[WEG2004\] 14h-001]{1}  & 14 16 49.44 & +52 25 30.8 & $18.55\pm0.02$ & $18.34\pm0.02$ & $18.10\pm0.02$ & $17.67\pm0.01$& $16.89\pm 0.03$ &\nodata\\
\nodata &\object[\[WEG2004\] 14h-118]{118}& 14 16 49.56 & +52 16 53.5 & $19.38\pm0.03$ & $19.73\pm0.03$ & $19.91\pm0.06$ & $20.13\pm0.06$& $17.87\pm 0.15$ &\nodata\\
\nodata &\object[\[WEG2004\] 14h-082]{82} & 14 16 49.90 & +52 18 09.1 & $19.10\pm0.02$ & $19.54\pm0.03$ & $19.62\pm0.04$ & $19.55\pm0.04$& $18.20\pm 0.10$ &\nodata\\
\nodata &\object[\[WEG2004\] 14h-056]{56} & 14 16 50.50 & +52 16 34.8 & $18.93\pm0.02$ & $19.10\pm0.02$ & $18.99\pm0.03$ & $18.53\pm0.02$& $17.05\pm 0.07$ &\nodata\\
\nodata &\object[\[WEG2004\] 14h-025]{25} & 14 16 50.92 & +52 15 28.5 & $18.93\pm0.02$ & $18.92\pm0.02$ & $18.80\pm0.03$ & $18.33\pm0.02$& $16.75\pm 0.06$ &\nodata\\
8     &\object[\[WEG2004\] 14h-009]{9}  & 14 16 51.27 & +52 20 46.0 & $18.42\pm0.02$ & $18.58\pm0.02$ & $18.52\pm0.02$ & $18.29\pm0.02$& $16.72\pm 0.03$ &\nodata\\
\object[\[NLA2005\] c009]{9}     &\nodata & 14 16 51.91 & +52 27 00.4 & $19.76\pm0.03$ & $20.04\pm0.04$ & $20.09\pm0.05$ & $19.99\pm0.05$& $17.85\pm 0.05$ &\nodata\\
\object[\[NLA2005\] c011]{11}    &\nodata & 14 16 53.74 & +52 21 24.2 & $19.05\pm0.02$ & $19.14\pm0.03$ & $19.14\pm0.03$ & $18.35\pm0.02$& $17.28\pm 0.09$ &\nodata\\
\nodata &\object[\[WEG2004\] 14h-047]{47} & 14 16 53.90 & +52 20 43.2 & $21.57\pm0.08$ & $21.33\pm0.07$ & $21.07\pm0.10$ & $20.71\pm0.08$& $>19.1        $ &\nodata\\
12    &\object[\[WEG2004\] 14h-054]{54} & 14 16 58.40 & +52 24 13.1 & $21.92\pm0.12$ & $21.73\pm0.11$ & $21.57\pm0.16$ & $20.91\pm0.10$& $19.00\pm 0.13$ &\nodata\\
\object[\[NLA2005\] c015]{15}    &\nodata & 14 17 00.14 & +52 23 04.2 & $19.61\pm0.03$ & $19.47\pm0.03$ & $19.67\pm0.04$ & $20.22\pm0.06$& $>19.1        $ &\nodata\\
16    &\object[\[WEG2004\] 14h-008]{8}  & 14 17 00.69 & +52 19 18.6 & $18.55\pm0.02$ & $18.56\pm0.02$ & $18.59\pm0.02$ & $18.20\pm0.02$& $16.89\pm 0.05$ &\nodata\\
18    &\object[\[WEG2004\] 14h-055]{55} & 14 17 04.21 & +52 24 53.6 & $18.16\pm0.02$ & $18.26\pm0.02$ & $18.54\pm0.02$ & $17.77\pm0.02$& $16.67\pm 0.06$ &\nodata\\
17    &\object[\[WEG2004\] 14h-022]{22} & 14 17 04.25 & +52 21 40.0 & $19.16\pm0.02$ & $19.36\pm0.03$ & $19.31\pm0.03$ & $18.92\pm0.03$& $17.88\pm 0.07$ &\nodata\\
\object[\[NLA2005\] c019]{19}    &\nodata & 14 17 05.68 & +52 31 46.8 & $21.09\pm0.06$ & $21.00\pm0.06$ & $21.38\pm0.13$ & $21.78\pm0.17$& $19.55\pm 0.19$ &\nodata\\
\object[\[NLA2005\] c020]{20}    &\nodata & 14 17 05.71 & +52 32 30.6 & $19.84\pm0.03$ & $19.62\pm0.03$ & $19.42\pm0.04$ & $19.03\pm0.03$& $17.37\pm 0.04$ &\nodata\\
\nodata &\object[\[WEG2004\] 14h-083]{83} & 14 17 06.89 & +52 18 08.9 & $19.67\pm0.03$ & $20.09\pm0.04$ & $20.02\pm0.06$ & $19.51\pm0.04$& $18.10\pm 0.21$ &\nodata\\
\object[\[NLA2005\] c021]{21}    &\nodata & 14 17 08.41 & +52 32 25.1 & $20.16\pm0.04$ & $20.10\pm0.04$ & $20.41\pm0.06$ & $20.75\pm0.08$& $19.24\pm 0.15$ &\nodata\\
22    &\object[\[WEG2004\] 14h-080]{80} & 14 17 08.70 & +52 29 30.1 & $21.32\pm0.07$ & $21.19\pm0.07$ & $21.37\pm0.13$ & $20.86\pm0.09$& $>19.1        $ &\nodata\\
23    &\object[\[WEG2004\] 14h-148]{148}& 14 17 08.89 & +52 27 09.1 & $18.60\pm0.02$ & $19.01\pm0.02$ & $19.37\pm0.03$ & $19.73\pm0.04$& $19.50\pm 0.16$ &\nodata\\
\nodata &\object[\[WEG2004\] 14h-030]{30} & 14 17 09.89 & +52 18 50.0 & $18.31\pm0.02$ & $18.32\pm0.02$ & $18.02\pm0.02$ & $17.29\pm0.01$& $16.21\pm 0.05$ &\nodata\\
25    &\object[\[WEG2004\] 14h-020]{20} & 14 17 10.56 & +52 28 29.0 & $19.22\pm0.03$ & $19.24\pm0.03$ & $19.39\pm0.03$ & $19.21\pm0.03$& $17.73\pm 0.05$ &\nodata\\
\object[\[NLA2005\] c026]{26}    &\nodata & 14 17 10.97 & +52 28 37.7 & $20.34\pm0.03$ & $20.27\pm0.03$ & $20.28\pm0.05$ & $19.85\pm0.04$& $18.68\pm 0.08$ &\nodata\\
27    &\object[\[WEG2004\] 14h-058]{58} & 14 17 11.06 & +52 25 40.8 & $18.48\pm0.02$ & $18.79\pm0.02$ & $19.23\pm0.03$ & $19.75\pm0.04$& $>19.1        $ &\nodata\\
\object[\[NLA2005\] c028]{28}    &\nodata & 14 17 11.61 & +52 31 32.5 & $19.80\pm0.03$ & $20.25\pm0.04$ & $20.54\pm0.07$ & $20.73\pm0.08$& $18.86\pm 0.11$ &\nodata\\
29    &\object[\[WEG2004\] 14h-017]{17} & 14 17 11.85 & +52 20 11.9 & $18.35\pm0.02$ & $18.44\pm0.02$ & $18.62\pm0.02$ & $18.10\pm0.02$& $16.75\pm 0.05$ &\nodata\\
\object[\[NLA2005\] c031]{31}    &\nodata & 14 17 14.30 & +52 25 33.3 & $20.97\pm0.06$ & $20.72\pm0.05$ & $20.67\pm0.07$ & $20.22\pm0.06$& $18.27\pm 0.08$ &\nodata\\
\object[\[NLA2005\] c0032]{32}    &\nodata & 14 17 14.83 & +52 34 19.3 & $20.00\pm0.04$ & $19.81\pm0.03$ & $19.87\pm0.05$ & $20.01\pm0.05$& $17.64\pm 0.07$ &\nodata\\
33    &\object[\[WEG2004\] 14h-029]{29} & 14 17 15.04 & +52 23 12.9 & $20.00\pm0.04$ & $19.72\pm0.03$ & $19.59\pm0.04$ & $19.11\pm0.03$& $18.04\pm 0.07$ &\nodata\\
\object[\[NLA2005\] c034]{34}    &\nodata & 14 17 15.21 & +52 26 50.5 & $19.23\pm0.03$ & $19.54\pm0.03$ & $19.73\pm0.04$ & $19.86\pm0.05$& $18.21\pm 0.08$ &\nodata\\
\nodata &\object[\[WEG2004\] 14h-144]{144}& 14 17 15.40 & +52 19 14.5 & $18.67\pm0.02$ & $18.86\pm0.02$ & $19.28\pm0.03$ & $18.98\pm0.03$& $18.25\pm 0.17$ &\nodata\\
\object[\[NLA2005\] c035]{35}    &\nodata & 14 17 18.96 & +52 27 44.2 & $20.46\pm0.05$ & $20.60\pm0.05$ & $20.94\pm0.08$ & $20.59\pm0.07$& $>19.1        $ &\nodata\\
\object[\[NLA2005\] c037]{37}    &\nodata & 14 17 19.28 & +52 27 55.9 & $19.58\pm0.03$ & $19.82\pm0.03$ & $20.38\pm0.06$ & $20.54\pm0.07$& $>19.1        $ &\nodata\\
38    &\object[\[WEG2004\] 14h-129]{129}& 14 17 20.05 & +52 25 00.5 & $21.10\pm0.06$ & $20.82\pm0.06$ & $20.17\pm0.05$ & $19.41\pm0.04$& $18.30\pm 0.13$ &\nodata\\
\object[\[NLA2005\] c039]{39}    &\nodata & 14 17 20.40 & +52 29 12.2 & $21.44\pm0.07$ & $21.28\pm0.07$ & $21.37\pm0.11$ & $21.54\pm0.14$& $19.24\pm 0.15$ &\nodata\\
\object[\[NLA2005\] c040]{40}    &\nodata & 14 17 22.95 & +52 31 43.7 & $19.66\pm0.03$ & $19.91\pm0.04$ & $20.16\pm0.05$ & $19.97\pm0.05$& $17.75\pm 0.23$ &\nodata\\
41    &\object[\[WEG2004\] 14h-028]{28} & 14 17 23.40 & +52 31 53.7 & $19.67\pm0.03$ & $19.70\pm0.03$ & $19.65\pm0.04$ & $19.23\pm0.03$& $17.60\pm 0.06$ &\nodata\\
\object[\[NLA2005\] c042]{42}    &\nodata & 14 17 23.60 & +52 25 55.1 & $19.58\pm0.03$ & $19.52\pm0.03$ & $19.47\pm0.03$ & $19.15\pm0.03$& $17.77\pm 0.06$ &\nodata\\
\object[\[NLA2005\] c043]{43}    &\nodata & 14 17 24.24 & +52 32 29.8 & $19.99\pm0.04$ & $20.31\pm0.04$ & $20.30\pm0.06$ & $20.34\pm0.06$& $17.68\pm 0.04$ &\nodata\\
44    &\object[\[WEG2004\] 14h-004]{4}  & 14 17 24.59 & +52 30 24.9 & $17.73\pm0.01$ & $17.58\pm0.01$ & $17.47\pm0.01$ & $17.20\pm0.01$& $16.33\pm 0.02$ &\nodata\\
\object[\[NLA2005\] c046]{46}    &\nodata & 14 17 25.27 & +52 35 44.3 & $21.65\pm0.08$ & $21.37\pm0.07$ & $20.87\pm0.09$ & $20.31\pm0.06$& $18.42\pm 0.08$ &\nodata\\
47    &\object[\[WEG2004\] 14h-114]{114}& 14 17 27.04 & +52 29 12.2 & $20.89\pm0.06$ & $21.02\pm0.06$ & $20.84\pm0.08$ & $20.50\pm0.07$& $19.49\pm 0.16$ &\nodata\\
\object[\[NLA2005\] c048]{48}    &\nodata & 14 17 27.28 & +52 31 31.5 & $21.16\pm0.06$ & $20.91\pm0.06$ & $21.09\pm0.09$ & $20.80\pm0.08$& $18.96\pm 0.11$ &\nodata\\
49    &\object[\[WEG2004\] 14h-149]{149}& 14 17 28.90 & +52 35 54.1 & $20.05\pm0.04$ & $20.02\pm0.04$ & $20.17\pm0.05$ & $20.33\pm0.06$& $18.64\pm 0.09$ &\nodata\\
\nodata &\object[\[WEG2004\] 14h-092]{92} & 14 17 29.88 & +52 37 16.4 & $20.12\pm0.04$ & $20.46\pm0.05$ & $20.78\pm0.09$ & $20.83\pm0.09$& $>19.1        $ &\nodata\\
50    &\object[\[WEG2004\] 14h-048]{48} & 14 17 29.95 & +52 27 47.9 & $20.72\pm0.05$ & $20.83\pm0.06$ & $20.98\pm0.09$ & $21.02\pm0.09$& $>19.1        $ &\nodata\\
\object[\[NLA2005\] c051]{51}    &\nodata & 14 17 30.59 & +52 22 43.4 & $21.04\pm0.06$ & $21.07\pm0.06$ & $21.72\pm0.16$ & $21.95\pm0.24$& $>19.1        $ &\nodata\\
\object[\[NLA2005\] c053]{53}    &\nodata & 14 17 30.71 & +52 23 04.8 & $22.01\pm0.10$ & $21.90\pm0.10$ & $21.93\pm0.19$ & $21.88\pm0.21$& $>19.1        $ &\nodata\\
\object[\[NLA2005\] c054]{54}    &\nodata & 14 17 30.81 & +52 28 18.6 & $20.83\pm0.05$ & $20.74\pm0.05$ & $20.73\pm0.07$ & $20.86\pm0.08$& $>19.1        $ &\nodata\\
55    &\object[\[WEG2004\] 14h-053]{53} & 14 17 32.62 & +52 32 03.2 & $19.16\pm0.03$ & $19.50\pm0.03$ & $19.85\pm0.04$ & $20.00\pm0.05$& $19.58\pm 0.19$ &15V10\\
57    &\object[\[WEG2004\] 14h-128]{128}& 14 17 33.82 & +52 33 49.4 & $19.27\pm0.03$ & $19.55\pm0.03$ & $19.67\pm0.04$ & $19.81\pm0.04$& $18.36\pm 0.06$ &\nodata\\
\object[\[NLA2005\] c058]{58}    &\nodata & 14 17 33.87 & +52 24 55.9 & $20.60\pm0.05$ & $20.63\pm0.05$ & $20.87\pm0.08$ & $21.01\pm0.10$& $18.32\pm 0.12$ &\nodata\\
\object[\[NLA2005\] c059]{59}    &\nodata & 14 17 34.37 & +52 31 06.9 & $18.48\pm0.02$ & $18.72\pm0.02$ & $19.27\pm0.03$ & $19.59\pm0.04$& $>19.1        $ &\nodata\\
60    &\object[\[WEG2004\] 14h-011]{11} & 14 17 34.85 & +52 28 10.6 & $18.53\pm0.02$ & $18.45\pm0.02$ & $18.45\pm0.02$ & $17.98\pm0.02$& $17.08\pm 0.03$ &\nodata\\
61    &\object[\[WEG2004\] 14h-002]{2}  & 14 17 35.95 & +52 30 29.8 & $18.75\pm0.02$ & $18.62\pm0.02$ & $18.40\pm0.02$ & $17.97\pm0.02$& $16.59\pm 0.03$ &15V12\\
\object[\[NLA2005\] c062]{62}    &\nodata & 14 17 36.30 & +52 30 16.8 & $22.33\pm0.11$ & $22.21\pm0.11$ & $22.73\pm0.30$ & $22.20\pm0.22$& $>19.1        $ &\nodata\\
\object[\[NLA2005\] c063]{63}    &\nodata & 14 17 36.33 & +52 35 44.4 & $20.20\pm0.04$ & $20.01\pm0.04$ & $19.73\pm0.04$ & $19.62\pm0.04$& $17.60\pm 0.08$ &\nodata\\
64    &\object[\[WEG2004\] 14h-045]{45} & 14 17 36.88 & +52 24 30.2 & $20.80\pm0.05$ & $20.74\pm0.05$ & $20.57\pm0.07$ & $20.46\pm0.07$& $>19.1        $ &\nodata\\
\nodata &\object[\[WEG2004\] 14h-153]{153}& 14 17 37.03 & +52 35 58.0 & $21.70\pm0.08$ & $21.98\pm0.10$ & $22.46\pm0.29$ & $22.06\pm0.20$& $>19.1        $ &\nodata\\
\object[\[NLA2005\] c065]{65}    &\nodata & 14 17 37.34 & +52 29 21.6 & $20.45\pm0.05$ & $20.45\pm0.05$ & $20.68\pm0.08$ & $20.68\pm0.08$& $19.32\pm 0.20$ &\nodata\\
66    &\object[\[WEG2004\] 14h-117]{117}& 14 17 38.68 & +52 34 14.0 & $21.03\pm0.06$ & $21.20\pm0.07$ & $21.52\pm0.13$ & $21.70\pm0.15$& $>19.1        $ &\nodata\\
67    &\object[\[WEG2004\] 14h-024]{24} & 14 17 38.86 & +52 23 33.1 & $19.89\pm0.04$ & $19.77\pm0.03$ & $19.52\pm0.04$ & $19.10\pm0.04$& $18.49\pm 0.25$ &\nodata\\
\object[\[NLA2005\] c068]{68}    &\nodata & 14 17 39.04 & +52 28 44.2 & $20.21\pm0.04$ & $20.19\pm0.04$ & $20.48\pm0.07$ & $20.57\pm0.08$& $19.09\pm 0.15$ &\nodata\\
\object[\[NLA2005\] c069]{69}    &\nodata & 14 17 39.30 & +52 28 50.3 & $19.35\pm0.03$ & $19.51\pm0.03$ & $19.95\pm0.05$ & $19.84\pm0.05$& $18.33\pm 0.08$ &\nodata\\
\object[\[NLA2005\] c070]{70}    &\nodata & 14 17 39.56 & +52 36 19.1 & $21.40\pm0.07$ & $21.34\pm0.07$ & $21.35\pm0.13$ & $21.05\pm0.10$& $>19.1        $ &\nodata\\
\object[\[NLA2005\] c071]{71}    &\nodata & 14 17 41.48 & +52 35 45.3 & $20.54\pm0.05$ & $20.46\pm0.05$ & $20.71\pm0.07$ & $20.82\pm0.08$& $>19.1        $ &\nodata\\
72    &\object[\[WEG2004\] 14h-023]{23} & 14 17 41.86 & +52 28 23.4 & $16.96\pm0.01$ & $16.48\pm0.01$ & $16.09\pm0.01$ & $15.60\pm0.01$& $14.51\pm 0.01$ &CUDSS14.13, 15V23\\
75    &\object[\[WEG2004\] 14h-122]{122}& 14 17 45.44 & +52 29 51.3 & $19.21\pm0.03$ & $19.27\pm0.03$ & $19.21\pm0.03$ & $18.62\pm0.02$& $17.17\pm 0.04$ &\nodata\\
76    &\object[\[WEG2004\] 14h-050]{50} & 14 17 45.62 & +52 28 02.0 & $20.50\pm0.05$ & $20.50\pm0.05$ & $20.48\pm0.07$ & $20.23\pm0.06$& $>19.1        $ &\nodata\\
77    &\object[\[WEG2004\] 14h-059]{59} & 14 17 45.95 & +52 30 32.6 & $18.92\pm0.02$ & $18.88\pm0.02$ & $18.79\pm0.02$ & $17.99\pm0.02$& $16.48\pm 0.02$ &15V28\\
\object[\[NLA2005\] c078]{78}    &\nodata & 14 17 46.13 & +52 25 27.2 & $19.71\pm0.03$ & $20.06\pm0.04$ & $20.24\pm0.06$ & $20.22\pm0.06$& $18.61\pm 0.22$ &\nodata\\
\object[\[NLA2005\] c079]{79}    &\nodata & 14 17 46.68 & +52 28 58.5 & $21.71\pm0.08$ & $21.59\pm0.08$ & $21.57\pm0.13$ & $22.09\pm0.22$& $>19.1        $ &\nodata\\
\object[\[NLA2005\] c080]{80}    &\nodata & 14 17 46.96 & +52 25 11.5 & $19.53\pm0.03$ & $20.11\pm0.04$ & $20.32\pm0.07$ & $20.86\pm0.11$& $18.51\pm 0.26$ &\nodata\\
\object[\[NLA2005\] c081]{81}    &\nodata & 14 17 47.04 & +52 28 16.8 & $23.17\pm0.19$ & $22.61\pm0.14$ & $21.97\pm0.20$ & $21.12\pm0.11$& $19.30\pm 0.20$ &\nodata\\
82    &\object[\[WEG2004\] 14h-102]{102}& 14 17 47.34 & +52 35 10.5 & $20.74\pm0.05$ & $20.59\pm0.05$ & $20.51\pm0.06$ & $20.11\pm0.05$& $19.08\pm 0.12$ &Westphal-MD106\\
\object[\[NLA2005\] c083]{83}    &\nodata & 14 17 49.21 & +52 28 03.4 & $18.77\pm0.02$ & $19.12\pm0.02$ & $19.51\pm0.04$ & $19.88\pm0.05$& $18.95\pm 0.16$ &15V34\\
84    &\object[\[WEG2004\] 14h-088]{88} & 14 17 49.23 & +52 28 11.7 & $19.68\pm0.03$ & $19.83\pm0.03$ & $20.04\pm0.05$ & $20.03\pm0.05$& $18.05\pm 0.08$ &15V33\\
\object[\[NLA2005\] c085]{85}    &\nodata & 14 17 49.80 & +52 31 44.3 & $21.08\pm0.06$ & $20.91\pm0.06$ & $20.84\pm0.08$ & $20.97\pm0.09$& $18.42\pm 0.07$ &\nodata\\
\object[\[NLA2005\] c086]{86}    &\nodata & 14 17 50.15 & +52 36 01.9 & $20.90\pm0.06$ & $20.78\pm0.05$ & $21.04\pm0.09$ & $20.85\pm0.08$& $19.24\pm 0.14$ &\nodata\\
\object[\[NLA2005\] c088]{88}    &\nodata & 14 17 50.87 & +52 36 33.5 & $21.12\pm0.06$ & $20.91\pm0.06$ & $20.54\pm0.06$ & $20.47\pm0.07$& $18.25\pm 0.10$ &\nodata\\
\object[\[NLA2005\] c089]{89}    &\nodata & 14 17 50.97 & +52 25 34.5 & $19.44\pm0.03$ & $19.57\pm0.03$ & $19.70\pm0.04$ & $19.57\pm0.05$& $>19.1        $ &\nodata\\
\object[\[NLA2005\] c091]{91}    &\nodata & 14 17 51.78 & +52 30 46.5 & $20.21\pm0.04$ & $20.52\pm0.05$ & $21.04\pm0.09$ & $20.93\pm0.09$& $19.18\pm 0.15$ &\nodata\\
\object[\[NLA2005\] c092]{92}    &\nodata & 14 17 52.41 & +52 28 53.2 & $19.93\pm0.04$ & $20.05\pm0.04$ & $20.39\pm0.06$ & $20.00\pm0.05$& $18.74\pm 0.13$ &\nodata\\
93    &\object[\[WEG2004\] 14h-133]{133}& 14 17 52.96 & +52 28 39.1 & $19.34\pm0.03$ & $19.79\pm0.03$ & $19.90\pm0.04$ & $20.27\pm0.06$& $18.46\pm 0.10$ &\nodata\\
\object[\[NLA2005\] c095]{95}    &\nodata & 14 17 53.68 & +52 34 46.2 & $19.55\pm0.03$ & $19.62\pm0.03$ & $19.47\pm0.04$ & $19.31\pm0.03$& $18.55\pm 0.08$ &\nodata\\
\object[\[NLA2005\] c096]{96}    &\nodata & 14 17 53.96 & +52 30 34.2 & $19.85\pm0.03$ & $19.81\pm0.03$ & $19.99\pm0.05$ & $19.31\pm0.03$& $18.10\pm 0.07$ &\nodata\\
\object[\[NLA2005\] c097]{97}    &\nodata & 14 17 54.24 & +52 31 23.5 & $19.77\pm0.03$ & $19.91\pm0.04$ & $20.14\pm0.05$ & $19.97\pm0.05$& $18.22\pm 0.06$ &\nodata\\
\object[\[NLA2005\] c098]{98}    &\nodata & 14 17 54.59 & +52 34 38.5 & $19.82\pm0.04$ & $20.17\pm0.04$ & $20.61\pm0.08$ & $20.75\pm0.09$& $>19.1        $ &\nodata\\
\object[\[NLA2005\] c099]{99}    &\nodata & 14 17 55.28 & +52 35 32.8 & $21.29\pm0.07$ & $21.04\pm0.06$ & $20.63\pm0.07$ & $19.98\pm0.05$& $18.77\pm 0.11$ &\nodata\\
101   &\object[\[WEG2004\] 14h-119]{119}& 14 17 56.84 & +52 31 24.7 & $20.52\pm0.05$ & $20.20\pm0.04$ & $20.14\pm0.05$ & $19.61\pm0.04$& $18.57\pm 0.28$ &\nodata\\
\object[\[NLA2005\] c102]{102}   &\nodata & 14 17 56.92 & +52 31 18.5 & $21.05\pm0.05$ & $20.99\pm0.05$ & $21.12\pm0.09$ & $21.00\pm0.09$& $19.32\pm 0.18$ &\nodata\\
103   &\object[\[WEG2004\] 14h-090]{90} & 14 17 57.10 & +52 26 31.3 & $19.37\pm0.03$ & $19.30\pm0.03$ & $19.64\pm0.05$ & $20.10\pm0.07$& $>19.1        $ &15V45\\
\object[\[NLA2005\] c104]{104}   &\nodata & 14 17 57.44 & +52 31 07.0 & $21.50\pm0.07$ & $21.36\pm0.07$ & $21.19\pm0.11$ & $20.60\pm0.08$& $18.75\pm 0.12$ &Westphal-M47\\
\object[\[NLA2005\] c106]{106}   &\nodata & 14 17 58.16 & +52 31 34.5 & $20.94\pm0.06$ & $20.89\pm0.06$ & $21.08\pm0.09$ & $21.45\pm0.13$& $>19.1        $ &15V47\\
108     &\object[\[WEG2004\] 14h-093]{93} & 14 17 58.94 & +52 31 39.1 & $19.73\pm0.03$ & $19.93\pm0.04$ & $19.97\pm0.05$ & $19.86\pm0.05$& $19.11\pm 0.15$ &\nodata\\
\object[\[NLA2005\] c112]{112}   &\nodata & 14 18 00.36 & +52 36 10.8 & $21.70\pm0.08$ & $21.60\pm0.08$ & $21.29\pm0.12$ & $20.93\pm0.09$& $>19.1        $ &\nodata\\
\object[\[NLA2005\] c111]{111}   &\nodata & 14 18 00.47 & +52 28 21.8 & $19.91\pm0.04$ & $19.82\pm0.03$ & $19.91\pm0.05$ & $19.79\pm0.05$& $17.26\pm 0.07$ &CUDSS14.3, 15V53\\
\object[\[NLA2005\] c113]{113}   &\nodata & 14 18 01.12 & +52 29 41.9 & $22.02\pm0.10$ & $21.32\pm0.07$ & $20.86\pm0.08$ & $19.70\pm0.04$& $18.32\pm 0.09$ &Westphal-oMD13\\
\object[\[NLA2005\] c114]{114}   &\nodata & 14 18 01.48 & +52 31 49.8 & $19.96\pm0.04$ & $20.25\pm0.04$ & $20.84\pm0.08$ & $20.89\pm0.09$& $>19.1        $ &\nodata\\
\object[\[NLA2005\] c115]{115}   &\nodata & 14 18 01.66 & +52 28 00.9 & $21.29\pm0.07$ & $21.15\pm0.07$ & $21.54\pm0.16$ & $21.47\pm0.19$& $>19.1        $ &\nodata\\
116   &\object[\[WEG2004\] 14h-052]{52} & 14 18 01.97 & +52 35 14.9 & $18.39\pm0.02$ & $17.88\pm0.01$ & $17.45\pm0.01$ & $16.98\pm0.01$& $15.97\pm 0.04$ &\nodata\\
\object[\[NLA2005\] c118]{118}   &\nodata & 14 18 02.86 & +52 35 47.5 & $21.47\pm0.08$ & $21.31\pm0.07$ & $21.07\pm0.10$ & $20.93\pm0.09$& $19.30\pm 0.18$ &\nodata\\
\object[\[NLA2005\] c119]{119}   &\nodata & 14 18 04.58 & +52 36 33.2 & $20.10\pm0.04$ & $19.86\pm0.04$ & $19.39\pm0.03$ & $18.78\pm0.03$& $17.63\pm 0.04$ &\nodata\\
\object[\[NLA2005\] c122]{122}   &\nodata & 14 18 06.48 & +52 33 58.2 & $19.46\pm0.03$ & $19.34\pm0.03$ & $19.72\pm0.04$ & $19.80\pm0.04$& $18.27\pm 0.08$ &\nodata\\
\object[\[NLA2005\] c124]{124}   &\nodata & 14 18 07.31 & +52 30 29.8 & $18.99\pm0.02$ & $19.33\pm0.03$ & $19.90\pm0.04$ & $20.17\pm0.06$& $>19.1        $ &\nodata\\
\object[\[NLA2005\] c126]{126}   &\nodata & 14 18 08.96 & +52 31 51.3 & $21.08\pm0.06$ & $20.85\pm0.06$ & $20.86\pm0.09$ & $20.74\pm0.09$& $18.82\pm 0.12$ &\nodata\\
\object[\[NLA2005\] c128]{128}   &\nodata & 14 18 11.22 & +52 30 12.4 & $20.88\pm0.06$ & $20.62\pm0.05$ & $20.58\pm0.07$ & $20.53\pm0.08$& $18.85\pm 0.19$ &Westphal-C50\\
\object[\[NLA2005\] c130]{130}   &\nodata & 14 18 13.07 & +52 31 12.9 & $19.62\pm0.02$ & $19.66\pm0.02$ & $19.84\pm0.04$ & $19.83\pm0.04$& $18.11\pm 0.10$ &15V72\\
\object[\[NLA2005\] c134]{134}   &\nodata & 14 18 15.38 & +52 32 47.7 & $20.90\pm0.06$ & $20.95\pm0.06$ & $21.01\pm0.09$ & $21.26\pm0.11$& $>19.1        $ &\nodata\\
\object[\[NLA2005\] c137]{137}   &\nodata & 14 18 16.29 & +52 33 31.1 & $20.18\pm0.04$ & $19.99\pm0.04$ & $20.23\pm0.05$ & $20.24\pm0.06$& $17.76\pm 0.07$ &\nodata\\
\object[\[NLA2005\] c139]{139}   &\nodata & 14 18 18.03 & +52 32 02.4 & $20.74\pm0.05$ & $20.68\pm0.05$ & $20.84\pm0.08$ & $21.02\pm0.11$& $>19.1        $ &\nodata\\
\object[\[NLA2005\] c141]{141}   &\nodata & 14 18 20.19 & +52 33 53.1 & $20.54\pm0.05$ & $20.32\pm0.04$ & $20.11\pm0.05$ & $19.93\pm0.05$& $>19.1        $ &\nodata\\
\object[\[NLA2005\] c143]{143}   &\nodata & 14 18 21.31 & +52 32 54.6 & $20.16\pm0.04$ & $19.95\pm0.04$ & $19.83\pm0.04$ & $19.63\pm0.04$& $17.71\pm 0.08$ &\nodata\\
\object[\[NLA2005\] c146]{146}   &\nodata & 14 18 22.53 & +52 36 07.2 & $20.28\pm0.04$ & $20.06\pm0.04$ & $20.07\pm0.05$ & $19.88\pm0.05$& $>19.1        $ &\nodata\\
\object[\[NLA2005\] c152]{152}   &\nodata & 14 18 26.41 & +52 32 36.0 & $16.75\pm0.01$ & $17.27\pm0.01$ & $17.70\pm0.01$ & $18.32\pm0.02$& $>19.1        $ &\nodata\\
\enddata
\tablecomments{The following X--ray sources are in the area covered by the IRAC and MIPS
images, but were not detected in the mid-infrared: Chandra IDs
\object[\[NLA2005\] c010]{10} (XMM ID 19), \object[\[NLA2005\] c013]{13} (XMM ID 84), \object[\[NLA2005\] c030]{30}, 
\object[\[NLA2005\] c036]{36}, \object[\[NLA2005\] c045]{45}, \object[\[NLA2005\] c052]{52} 
and XMM IDs \object[\[WEG2004\] 14h-060]{60}, \object[\[WEG2004\] 14h-064]{64}, \object[\[WEG2004\] 14h-078]{78}, 
\object[\[WEG2004\] 14h-081]{81}, \object[\[WEG2004\] 14h-096]{96}, \object[\[WEG2004\] 14h-121]{121}, 
\object[\[WEG2004\] 14h-123]{123}.}
\tablenotetext{a}{Source catalog number given in Table 3 of \citet{egs_cxo}, referred to in this paper as cNNN. Official source
designation as given in NED is [NLA 2005] NNN.}
\tablenotetext{b}{Source number in Table 2 of \citet{was04}, referred to in this paper as xNNN. Official source
designation as given in NED is [WEG 2004] 14h-NNN.}
\tablenotetext{c}{15V: \citet{egs_vla91}, CUDSS: \citet{webb03}, Westphal: \citet{ccs03}}
\end{deluxetable}

\end{document}